\documentclass[%
 aip,
 amsmath,amssymb,
 reprint,
]{revtex4-1}

\usepackage{graphicx}
\usepackage{dcolumn}
\usepackage{bm}

\usepackage[utf8]{inputenc}
\usepackage[T1]{fontenc}
\usepackage{mathptmx}
\usepackage{etoolbox}

\makeatletter
\def\@email#1#2{%
 \endgroup
 \patchcmd{\titleblock@produce}
  {\frontmatter@RRAPformat}
  {\frontmatter@RRAPformat{\produce@RRAP{*#1\href{mailto:#2}{#2}}}\frontmatter@RRAPformat}
  {}{}
}%
\makeatother
\begin{document}

\preprint{AIP/123-QED}

\title{Enhancing search efficiency through diffusive echo}

\author{Charles Antoine}
 \email{charles.antoine@sorbonne-universite.fr}
 
\affiliation{
	Laboratoire de Physique Th\'eorique de la Mati\`ere Condens\'ee, UPMC, CNRS  UMR 7600, Sorbonne Universit\'es, 4, place Jussieu, 75252 Paris Cedex 05, France
}
\author{Julian Talbot}
 \email{talbot@lptmc.jussieu.fr}
 
\affiliation{
	Laboratoire de Physique Th\'eorique de la Mati\`ere Condens\'ee, UPMC, CNRS  UMR 7600, Sorbonne Universit\'es, 4, place Jussieu, 75252 Paris Cedex 05, France
}

 

\date{\today}

\begin{abstract}
Despite having been studied for decades, first passage processes remain an active area of research. In this contribution we examine a  particle diffusing in an annulus with an inner absorbing boundary and an outer reflective boundary. We obtain analytic expressions for the joint distribution of the hitting time and the hitting angle in two and three dimensions. For certain configurations we observe a ``diffusive echo", i.e. two well-defined maxima in the first passage time distribution to a targeted position on the absorbing boundary. This effect, which results from the interplay between the starting location and the environmental constraints, may help to increase the efficiency of the random search by generating a sustained flux to the targeted position over a short period. Finally, we examine the corresponding one-dimensional system for which there is no well-defined echo.
\end{abstract}

\maketitle


\section{Introduction}

\begin{figure}[t]
\begin{center}
\resizebox{7cm}{!}{\includegraphics{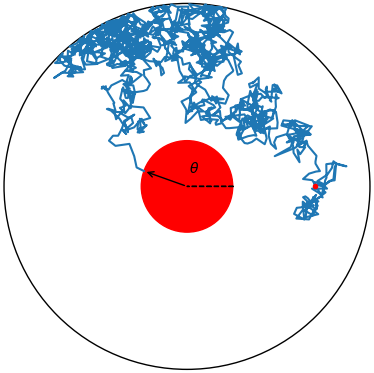}}
\end{center}
\caption{The 2D system consists of a Brownian particle diffusing in an annulus formed from an inner absorbing disk of radius $a=1$ and an outer reflecting wall of radius $b$ (here equal to $4$ for illustration). 
The diffusive particle trajectory starts at a distance $r_0$ from the center (small red point) and terminates when it touches the inner disk.
The targeted position is labeled by the angle $\theta$, while the starting position corresponds to $\theta = 0$.}\label{fig:system}
\end{figure}

What is the probability that a diffusing agent arrives 'alive' at a specific target inside a highly constrained and hazardous environment?
This question is pertinent in a wide range of fields from cellular biology and chemical reactions, to materials science, animal foraging and spreading of viruses \cite{Qiu2017,Bell2017,Okubo2001,viswanathan2011physics,Mizumoto2018,Mohapatra2019,Nayak2020Predators,Lloyd2001viruses,Wang2021airborne,LuLu2021SARSCoV2}. 
Examples include DNA and mRNA polymers diffusing passively inside the cell cytoplasm with the task of safely finding a nuclear pore to enter the cell nucleus \cite{Zhang2016,Kar2018,Schavemaker2018ImportantDiffusion,erbas2019DNA,stracy2021DNAtarget}. Or, more generally, for any brownian search process in a confined environment where traps or lethal zones may terminate the particle trajectory \cite{redner2001guide,Bray2013MajumdarSchehr,Bressloff2013,Metzler2014first,benichou2014first,Godec2016}.

The outcome depends on many factors including the system geometry, the boundary conditions, the shape and mobility of the searcher and the target, the number and starting position of the searching agents, and the degree of hindrance due to crowding and possible traps. 
If the targeted site is behind an absorbing obstacle, one may wonder where to start the diffusive search process in order to optimize its efficiency. Too close to the obstacle, and the brownian trajectory will likely be killed before reaching its destination. Too far, and the diffusive exploration lengthens in a prohibitive way.

The presence of a boundary may radically affect the search, whether or not it is diffusive. In the context of sound propagation for example, a confining wall may reflect a distant voice, creating one or multiple echos.

For diffusive search processes, one no longer has a sound-like propagation but the question of a `diffusive echo' effect is open. If it exists - and we show it does - one may wonder how it depends on the various parameters of the system and how it may enhance search efficiency. Of particular interest is the question of its persistence for a strong confinement in one, two and three dimensions.

The term 'diffusive echo' was first introduced in 1994 for steady two-dimensional flows \cite{koplik1994tracer,koplik1995universal}. The effect is, however, tenuous, consisting of a subtle alteration of the transit time distribution at long times. More specifically, a shoulder is present between the usual power law decay at short times and the confinement-induced exponential decay at long times. This, however, cannot be considered as a true echo phenomenon, like the acoustic one, where two or more well defined maxima appear successively.

Geometric confinement is known to play a crucial role in first passage processes \cite{Condamin2007nature,Benichou2010NatChem,VAT2015,Benichou2015JPhysA,Godec2016a,Grebenkov2016prl}, be the external boundary totally or partially reflecting \cite{Grebenkov2010pre,Guerin2021ImperfectReactions}, with narrow \cite{schuss2007narrow,Holcman2014,Grebenkov2019narrow} or large escape zones and/or absorbing areas, part of which may play the role of target. Among notable recent applications of such confined processes, we can cite intracellular targeting for therapeutic intervention on oncogenic and degenerative diseases \cite{Hong2013,Yang2014,2016BoundaryIssues,Ozay2016,Miersch2016,Trenevska2017,Lanoisele2018,Sil2020CD44,Taylor2020ProteinTracking}.

Diffusion phenomena, including heat transfer, are described theoretically by solving differential equations with the appropriate boundary conditions. The methodology, as well as solutions for many simple models, were established a long time ago \cite{carslaw1959conduction,crank1975mathematics,holman1990heat}, and underlie ongoing research \cite{Bickel2007,Benichou2010NatChem,Grebenkov2010pre,VAT2015,Godec2016a,Grebenkov2016prl,antoine2020nonintuitive,Guerin2021ImperfectReactions,grebenkov2019spectral,grebenkov2020diffusion,grebenkov2020surface,grebenkov2020paradigm}.

In this article we consider the model shown in Fig. \ref{fig:system} where a spherical absorbing surface of radius $a$ is contained within a spherically symmetric reflecting wall of radius $b$. The trajectory of a diffusing particle ends the instant it touches the inner, absorbing surface (perfect reactivity).
In this work we are concerned with the joint probability that the diffusing particle hits a specific position on the central absorbing surface at time $t$. 
We first consider the two-dimensional case  and then compare with the corresponding one and three dimensional systems.

\section{Time dependent hitting distribution in two dimensions}

We seek the time-dependent hitting distribution at a specific position on the inner absorbing surface. This quantity, also known as first passage time distribution, may be obtained from the diffusive flux, which is proportional to the gradient of the particle concentration.

For the two dimensional problem, the time-dependent density evolves according to
\begin{equation}\label{eq:DE}
D\left(\frac{1}{r}\frac{\partial}{\partial r}\left(r\frac{\partial c}{\partial r}\right)+\frac{1}{r^2}\frac{\partial^2 c}{\partial \theta^2}\right)
=\frac{\partial c}{\partial t}
\end{equation}
with absorbing and reflecting boundaries at $r=a$ and $r=b$, respectively:
\begin{eqnarray}\label{eq:bcs2D}
c(a,\theta,t)&=0,\nonumber\\
\left.\frac{\partial c}{\partial r}\right|_{r=b}&=0,
\end{eqnarray}
and the initial condition
\begin{equation}\label{eq:delta2D}
c(r,\theta,0)=\frac{1}{r}\delta(r-r_0)\delta(\theta-\theta_0)
\end{equation}
The diffusion coefficient $D$ is assumed to be homogeneous and constant. The flux on a $d \theta$ wedge of the target is simply equal to $a d \theta D\left.\frac{\partial c}{\partial r}\right|_{r=a} $, leading to the time-dependent hitting distribution
\begin{equation}\label{eq:hittinganguleq}
j(\theta,t)=a\;D\left.\frac{\partial c}{\partial r}\right|_{r=a}
\end{equation}
where the coordinate $\theta$ labels the targeted position on the inner circle, while the starting position corresponds to $\theta_0 = 0$.

To solve Eq. (\ref{eq:DE}) we assume a solution of the form
$c(r,\theta,t)=R(r)\Theta(\theta)T(t)$. 
This leads to an ordinary differential equation for the radial function
\begin{equation}
R''+\frac{1}{r}R'+\left(k^2-\frac{m^2}{r^2}\right)R=0
\end{equation}
which can be expressed in terms of Bessel functions.

As detailed in Appendix~\ref{app:two}, the full solution of Eq. (\ref{eq:DE}), which is periodic in $\theta$ and which satisfies the boundary conditions Eq. (\ref{eq:bcs2D}), is
\begin{equation}\label{eq:nrthetat}
c(r,\theta,t)=\frac{\pi}{2} \sum_{m=-\infty}^{\infty} \sum_{n=1}^{\infty} \cos(m\theta) \frac{F_{mn}(r) F_{mn}(r_0)}{C_{mn}} k_{mn}^2 e^{-k_{mn}^2Dt}
\end{equation}
with the $F_{mn}$ function defined as
\begin{equation}
F_{mn}(r)=Y_m(k_{mn}r)J_m(k_{mn}a)-J_m(k_{mn}r)Y_m(k_{mn}a)
\end{equation}
where $J_m(x)$ and $Y_m(x)$ are the Bessel functions of the first and
second kind, respectively, and $k_{mn}$ is the $n$th root of the auxiliary equation 
\begin{eqnarray}\label{eq:rootk}
J_m(ka)(Y_{m-1}(kb)-Y_{m+1}(kb))\nonumber\\
= Y_m(ka)(J_{m-1}(kb)-J_{m+1}(kb))
\end{eqnarray}
for $m \geq 1$ and
\begin{equation}\label{eq:rootkm0}
J_0(ka)Y_{1}(kb) - Y_0(ka)J_{1}(kb) = 0
\end{equation}
for $m=0$.
These roots do not depend on the initial position $r_0$ and depend solely on $m$ and the boundary positions $a$ and $b$.
The expression for the coefficients $C_{mn}$ is given in the Appendix~\ref{app:two}.
Note that Eq. (\ref{eq:nrthetat}) is symmetric in $r$ and $r_0$, and depends on these two positions only through the function $F_{mn}$.

The instantaneous flux at time $t$ on the target at the position labeled by the angle $\theta$ is obtained directly from Eq. (\ref{eq:hittinganguleq}): 
\begin{equation}\label{eq:flux2Dmain}
j(\theta,t) = D \sum_{m=-\infty}^{\infty} \sum_{n=1}^{\infty}  \frac{\cos(m\theta) F_{mn}(r_0) k_{mn}^2 e^{-k_{mn}^2Dt}}{C_{mn}}
\end{equation}

The expressions for the density and hitting distribution can be compared with those of the corresponding unbounded system ($b=\infty$, indicated by an asterisk) \cite{carslaw1959conduction}:
\begin{equation}\label{eq:nrthetatunbounded}
c^{*}(r,\theta,t)= \sum_{m=-\infty}^{\infty} \frac{\cos(m\theta)}{2\pi} \int_{0}^{\infty} \frac{F_{m}(r) F_{m}(r_0) k e^{-k^2Dt} dk}{J_m^2(ka) + Y_m^2(ka)}
\end{equation}
and 
\begin{equation}\label{eq:epsitildeunbounded}
j^{*}(\theta,t) = \frac{D}{\pi^2} \sum_{m=-\infty}^{\infty} \cos(m\theta) \int_{0}^{\infty}  \frac{ F_{m}(r_0) k e^{-k^2Dt} dk}{J_m^2(ka) + Y_m^2(ka)}
\end{equation}
where $F_m$ is
\begin{equation}
F_{m}(r)=Y_m(kr)J_m(ka)-J_m(kr)Y_m(ka)
\end{equation}

As shown in \cite{antoine2020nonintuitive}, the eventual hitting angle distribution $\tilde{J}(\theta)=\int_0^{\infty} j(\theta,t)dt$
for the bounded case is given by
\begin{equation}\label{eq:bound2D}
\tilde{J}(\theta)=\frac{1}{2\pi}\left[1+2\sum_{m=1}^{\infty}c_m\cos(m\theta)\right]
\end{equation}
with
\begin{equation}
c_m=\frac{r_0^{2m}+b^{2m}}{a^{2m}+b^{2m}}\left(\frac{a}{r_0}\right)^m
\end{equation}
which reduces to the unbounded case (see e.g. \cite{redner2001guide})in the limit $b \rightarrow \infty$
\begin{equation}\label{eq:unbound2D}
\tilde{J}^{*}(\theta)=\frac{1}{2\pi}\frac{1-a^2/r_0^2}{1-(2a/r_0)\cos\theta+a^2/r_0^2},
\end{equation}

The distribution (\ref{eq:bound2D}) is normalized, $\int_{-\pi}^{\pi}d\theta \tilde{J}(\theta)=1$, which is consistent with the certainty of (eventually) hitting the target in two dimensions (Pólya's recurrence theorem \cite{Polya1921,redner2001guide}).

\section{Diffusive echo in 2D}

For all values of $\theta$ and $r_0 \in ]a,b]$, the instantaneous local target flux $j(\theta,t)$ initially increases from zero, then passes through one or two maxima, before decreasing exponentially to zero. The short-time evolution of $j(\theta,t)$ is similar to $j^{*}(\theta,t)$, but the bounded system may display two maxima at intermediate times, an effect referred to as a 'diffusive echo' in the following.

\begin{figure}[h]
	\begin{center}
		\resizebox{8.5cm}{!}{\includegraphics{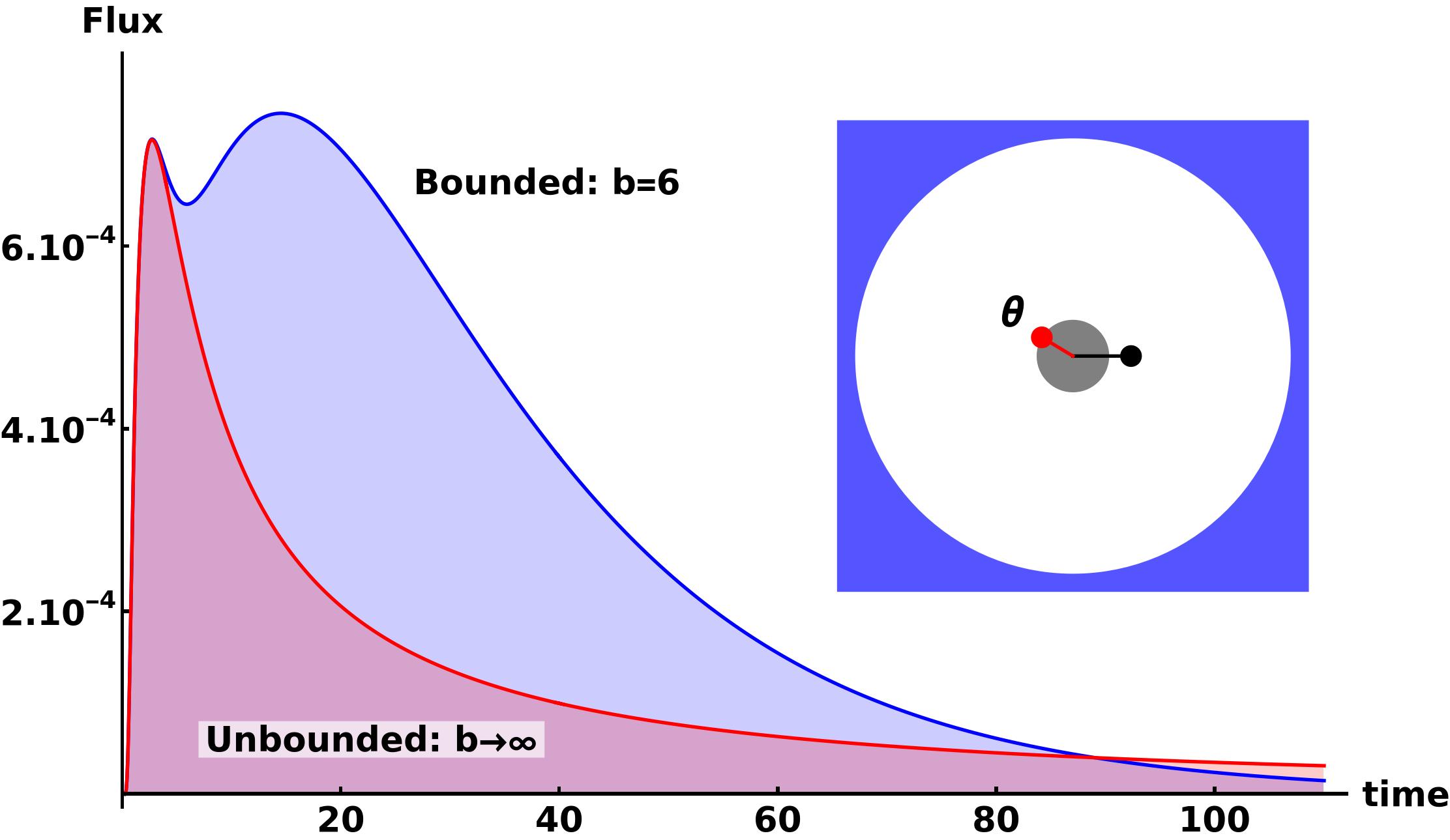}}
	\end{center}
	\caption{Manifestation of a diffusive echo. Time evolution of the hitting angle distribution $j(\theta,t)$ of a bounded system ($b=6$, upper curve) compared with  $j^{*}(\theta,t)$ for an unbounded system ($b=\infty$, lower curve). In both cases: $a=1$, $r_0=1.6$ and $\theta=5\pi/6$.} \label{fig:Figure-Bounded-b6-vs-Unbounded}
\end{figure}

An example of a double maximum is shown in Fig. \ref{fig:Figure-Bounded-b6-vs-Unbounded} for $a=1$, $b=6$, $r_0=1.6$, and for a particular hitting angle $\theta=5\pi/6$. The unbounded distribution is also shown for comparison. As expected, the instantaneous flux to any target position the targeted position is higher for a bounded system until about $t\approx 90$ ($2.5$ in units of $b^2/D$) since, in the unbounded case, part of the flux leaks into the external region $r>b$. For times greater than this, the opposite trend occurs. The slow decrease in the unbounded case is due to the possibility for the diffusing particle to return to the target after a long time away from it, whereas in the bounded case the particle is consumed rapidly.

\begin{figure}[h]
	\begin{center}
		\resizebox{9cm}{!}{\includegraphics{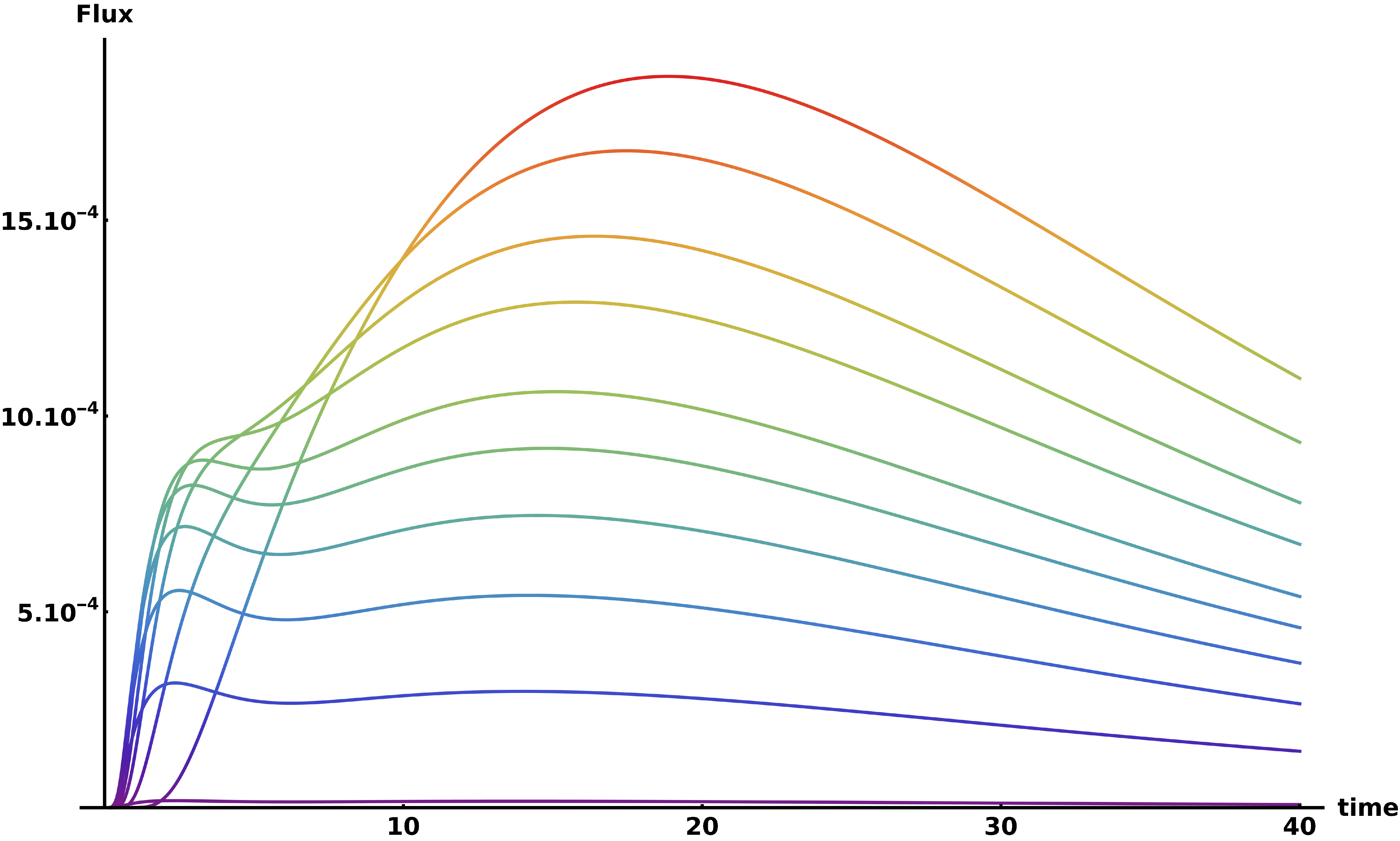}}
	\end{center}
	\caption{Flux $j(\theta,t)$ at a hitting angle $\theta=5\pi/6$ for various starting points. From top to bottom $r_0=$ $6$, $3.6$, $2.8$, $2.4$, $2$, $1.8$, $1.6$, $1.4$, $1.2$, $1.01$. The confinement is $b=6$.}\label{fig:Figure-Bounded-b6-vs-variousr0}
\end{figure}

While the first peak of $j(\theta,t)$ is common to the bounded and unbounded cases, the presence of the confining boundary leads to a second peak in the target local flux. A series of such curves for different values of the initial position between $a < r_0 < b$ is shown in Fig. \ref{fig:Figure-Bounded-b6-vs-variousr0}.

We observe a switching between two diffusion regimes. For $r_0$ far from the target and close to the reflecting boundary, $j(\theta,t)$ is totally different from the unbounded case and is strongly influenced by the confinement. On the other hand, for $r_0$ far enough from the external boundary, the latter is not felt at short times and the flux to the target is controlled by ``direct" diffusion. At later times, an indirect flux, due to reflection by the external boundary, is dominant. For intermediate values of $r_0$, these two regimes superpose and a double maximum in $j(\theta,t)$ may appear.

An interesting consequence of the double maximuma of Figs. \ref{fig:Figure-Bounded-b6-vs-Unbounded} and \ref{fig:Figure-Bounded-b6-vs-variousr0} is that the flux is sustained at a high value for a much longer time than in the unbounded case. This  can be quantified by comparing the length of time during which the hitting distribution is higher than a specific value, for both the bounded and unbounded cases. This specific threshold value may be chosen as the value of the local minimum of $j(\theta,t)$, i.e. between its two consecutive maxima. For example, for $r_0 = 1.6$, one gets a flux of $6.5$ $10^{-4}$ (in units of $D/a^2$) as shown in Fig. \ref{fig:Figure-Bounded-b6-vs-Unbounded}. For this flux value, the width of the distribution peak of the bounded system is more than $8.5$ times the width of the unbounded distribution. For $r_0 = 2$, the difference is even larger: the bounded distribution peak is more than $15$ times larger than the unbounded distribution.

\begin{figure}[h!]
	\begin{center}
		\resizebox{8.5cm}{!}{\includegraphics{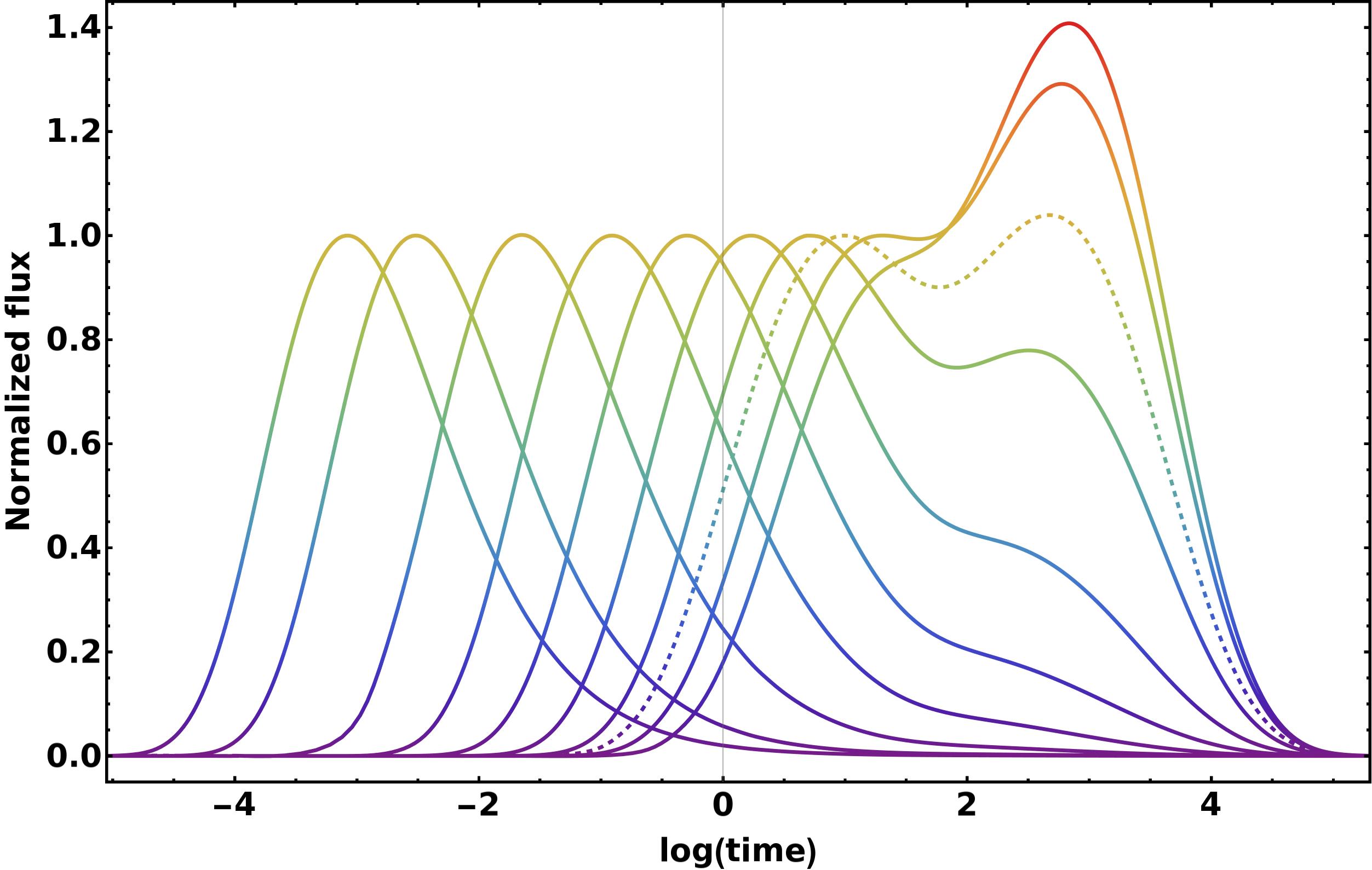}}
	\end{center}
	\caption{Normalized flux $j\left( \theta,t \right) / j_\textrm{max}$ at different hitting angles $\theta$ from the same starting point $r_0 = 1.6$. From left to right $\theta =$ $0$, $0.4$, $0.8$, $1.2$, $1.6$, $2$, $2.4$, $2.8$, $\pi$. The case $\theta = 5\pi/6$ of Fig. \ref{fig:Figure-Bounded-b6-vs-Unbounded} is also shown for comparison (dashed line). The confinement is $b=6$.}\label{fig:Figure-Bounded-beq6-r0-1p6-vs-various-Theta0-NormalizedFlux-log-of-time}
\end{figure}

As already underlined, the confining boundary leads to a redistribution of the hitting probability in the course of time. The time integral of $j(\theta,t)$ can therefore approach its infinite time value long before the time integral of $j^*(\theta,t)$. In other words, the probability to reach a specific target can be much more concentrated for a confined system than for an unbounded one, providing the starting position $r_0$ is well chosen.
The diffusive echo phenomenon is thus not merely a mathematical curiosity, but may be exploited to significantly enhance the efficiency of a passive diffusive search. More specifically, for example, it may play a role in determining \textit{where} a therapeutic molecule should  be injected in a cell in order to more efficiently reach a targeted zone in the inner core of the cell. For a core of one-sixth of the cell size, $b=6a$, the ideal starting location is close to $r_0 \simeq 2a$ for a target located at the angle $\theta = 5\pi/6$, i.e. on the opposite side of the core from the departure point.

The dependence of the diffusive echo on $\theta$ is subtle (see Fig. \ref{fig:Figure-Bounded-beq6-r0-1p6-vs-various-Theta0-NormalizedFlux-log-of-time}).
For small and large $\theta$, only one maximum is present, i.e. there is no diffusive echo. The explanation relies on the sink effect of the absorbing inner circle. For small values of $\theta$, the direct flux to the targeted point specified by $\theta$ is not impeded by the remaining part of the absorbing inner circle. As a consequence, the direct flux to the target is sufficiently high so that it is not overcome by the subsequent indirect flux from the outer boundary. On the other hand, for a targeted point on the opposite side of the inner circle ($\theta \simeq \pi$), the absorbing inner boundary plays the role of a sink that prevents direct diffusion towards the targeted point, allowing the confinement-controlled diffusion to dominate. For intermediate $\theta$ values, these two processes compete and a double maximum in  $j(\theta,t)$ may appear. On the right hand side of Fig. \ref{fig:Figure-Bounded-beq6-r0-1p6-vs-various-Theta0-NormalizedFlux-log-of-time}, one clearly sees the gradual build-up of a second, reflection-induced flux maximum. For $\theta=5\pi/6$ (dashed line in Fig. \ref{fig:Figure-Bounded-beq6-r0-1p6-vs-various-Theta0-NormalizedFlux-log-of-time}), the diffusive echo is more apparent since the two maxima have a similar amplitude.

\begin{figure}[h]
	\begin{center}
		\resizebox{8.5cm}{!}{\includegraphics{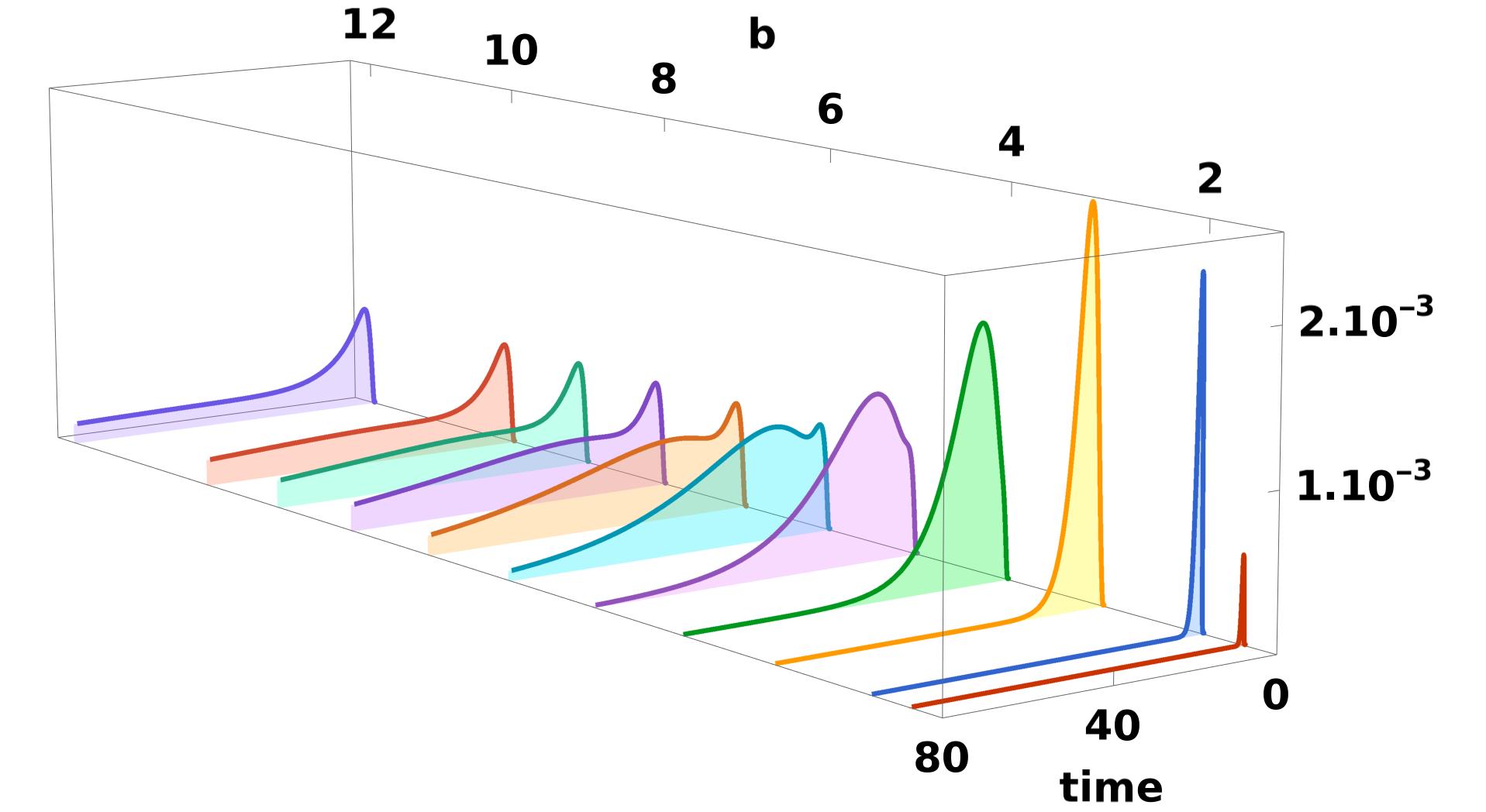}}
	\end{center}
	\caption{Flux $j(\theta,t)$ at a hitting angle $\theta=5\pi/6$ from the same starting point $r_0 = 1.6$ but for various values of the confining parameter $b$. From bottom right to top left: $b=$ $1.6$ (coinciding with the starting point), $2$, $3$, $4$, $5$, $6$, $7$, $8$, $9$, $10$, $12$. }\label{fig:Figure-Bounded-r0eq1p6-Theta0-2p6-vs-various-b-ListLinePlot3D-v2}
\end{figure}

\begin{figure}[h]
	\begin{center}
		\resizebox{8.5cm}{!}{\includegraphics{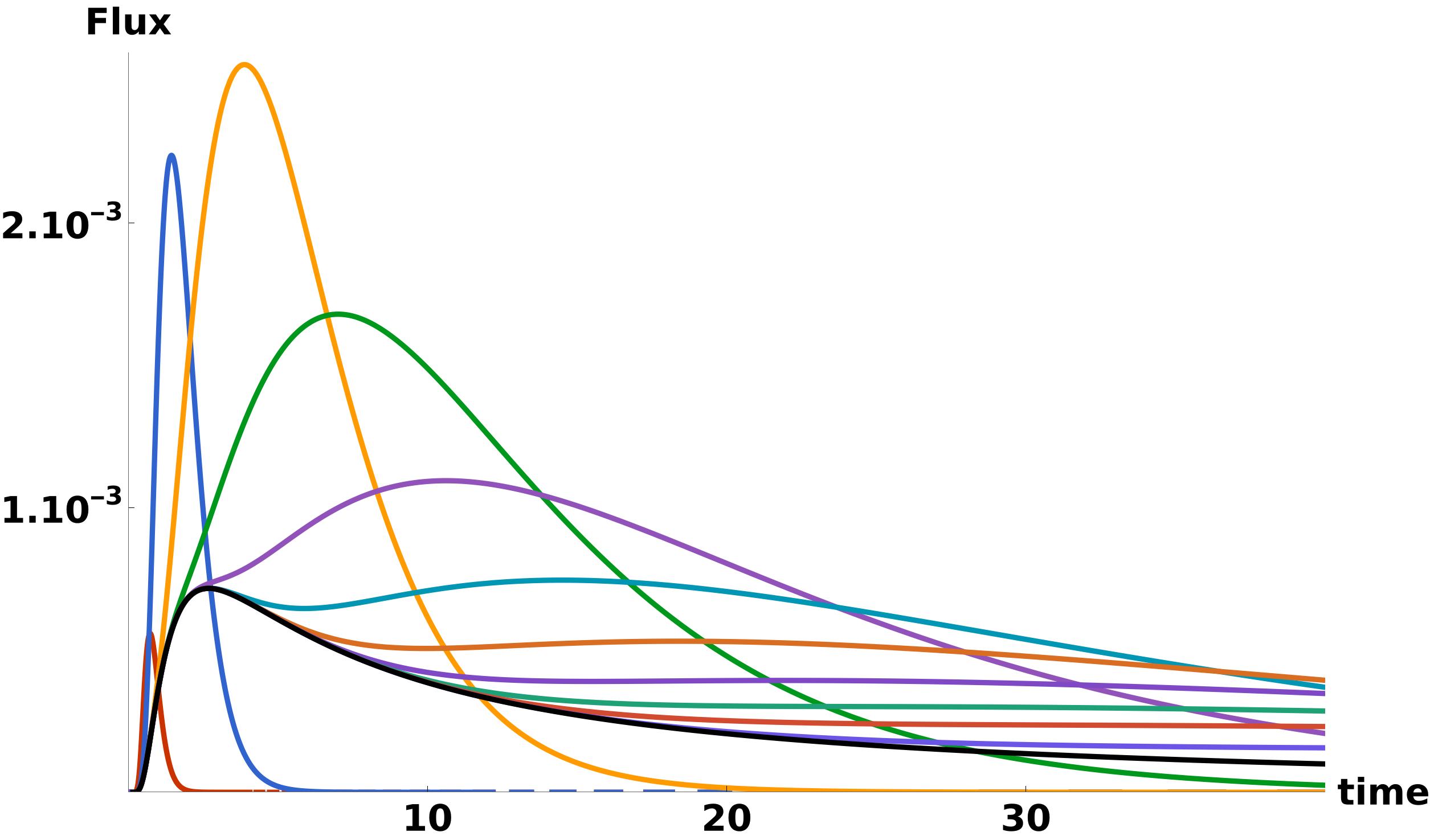}}
	\end{center}
	\caption{Same curves as in Fig. \ref{fig:Figure-Bounded-r0eq1p6-Theta0-2p6-vs-various-b-ListLinePlot3D-v2} (with the same color code), but regrouped in a single 2D graphic. The additional black curve corresponds to the unbounded case ($b \rightarrow \infty$).}\label{fig:Figure-Bounded-r0eq1p6-Theta0-2p6-vs-various-b-v2}
\end{figure}

The dependence of the diffusive echo on the degree of confinement may be explained similarly (see Fig. \ref{fig:Figure-Bounded-r0eq1p6-Theta0-2p6-vs-various-b-ListLinePlot3D-v2} and \ref{fig:Figure-Bounded-r0eq1p6-Theta0-2p6-vs-various-b-v2}). For small values of $b$, the reflecting boundary guides the diffusing particle around the central circle. Due to the proximity of the absorbing inner boundary, there is almost no direct diffusion toward the specific targeted $\theta$-point, particularly if it is on the opposite side of the starting point. All possible 'direct' trajectories are indeed too close to the absorbing central zone and only 'indirect' trajectories can survive and contribute to the flux on the targeted point, especially when the starting point is close to the inner absorbing circle (i.e. for small values of $r_0$).
For larger values of $b$, i.e. for a weaker confinement, this indirect flux decreases since many of the indirect trajectories seep from the target area and are no longer focused on it. The intensity of the direct flux becomes comparable with the reflected one, enabling a diffusive echo on the targeted point since these two fluxes are time shifted. Finally, for very weak confinement ($b \gg a$), the reflected flux is tenuous and one recovers the unbounded behavior.

 For very strong confinement, we observe a narrow distribution of the flux whose maximum occurs at a much shorter time than in the unbounded case (see Fig. \ref{fig:Figure-Bounded-r0eq1p6-Theta0-2p6-vs-various-b-v2}, the first two curves on the left). This effect strongly depends on the starting position $r_0$ as expected from a simple explanation in terms of survival of trajectories. Indeed, for $r_0$ very close to the confining boundary, almost all the trajectories are focused initially towards the absorbing central zone. For highly confined systems (small $b$), these trajectories cannot escape and are absorbed due to the lack of space. As a consequence, the average time needed to reach a specific targeted point is reduced since almost no trajectories can survive for a long time. On the contrary, as soon as $r_0$ is no longer close to $b$, some trajectories may escape through the back side (or on the left/right sides) during the first instants of the diffusion process and  then be guided by the confining boundary, leading to a higher contribution to the flux on a distant point. Among all trajectories, the weight of the longest ones increase in the distribution of the hitting times: the flux maximum is thus higher than in the unbounded case, and occurs at a later time. Then, as explained above, for higher values of $b$, the space available to the diffusing particle becomes sufficient to allow for longer and longer survival times, and the hitting times distribution broadens and flattens. There is therefore an optimum value of the confining parameter $b$ for which the flux on a specific targeted point is the highest (this optimal $b$ value being close to $3$ for $r_0 = 1.6$ as shown in Fig. \ref{fig:Figure-Bounded-r0eq1p6-Theta0-2p6-vs-various-b-v2}).

\begin{figure}[htp]
	
	\centering

\includegraphics[width=60mm]{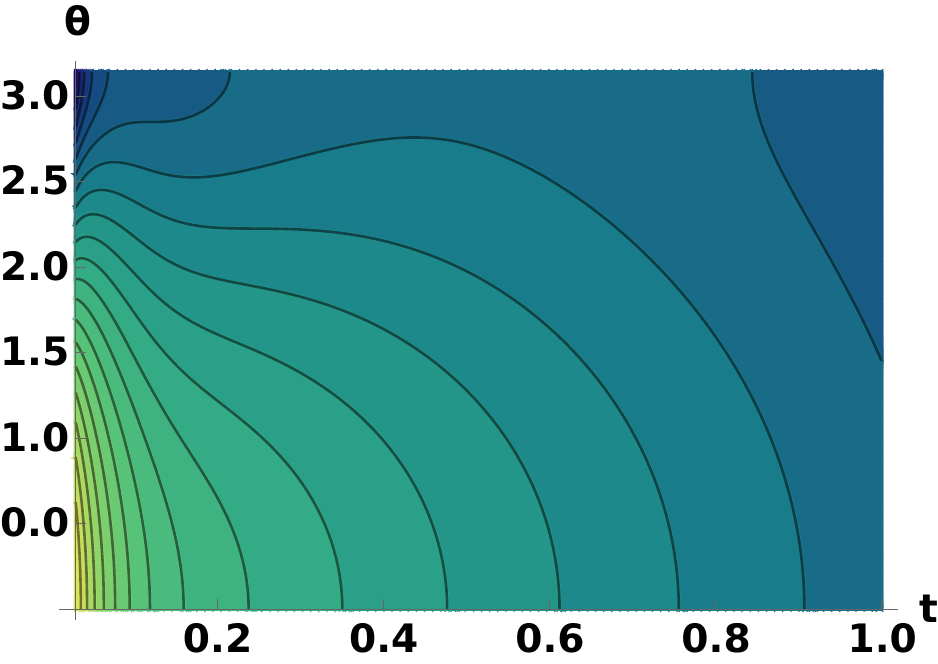}\\
    
\includegraphics[width=60mm]{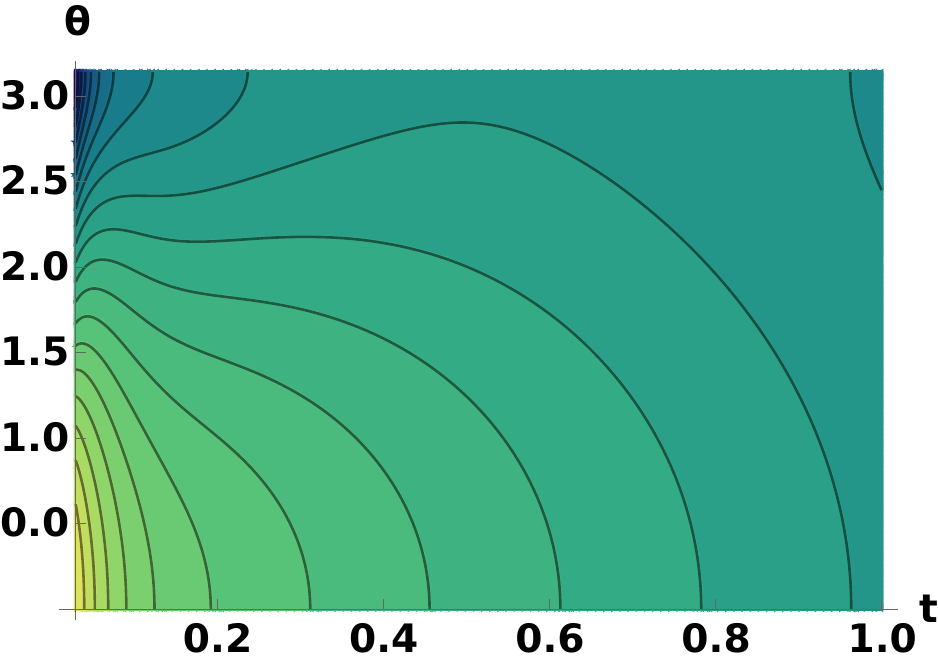}\\
	
\includegraphics[width=60mm]{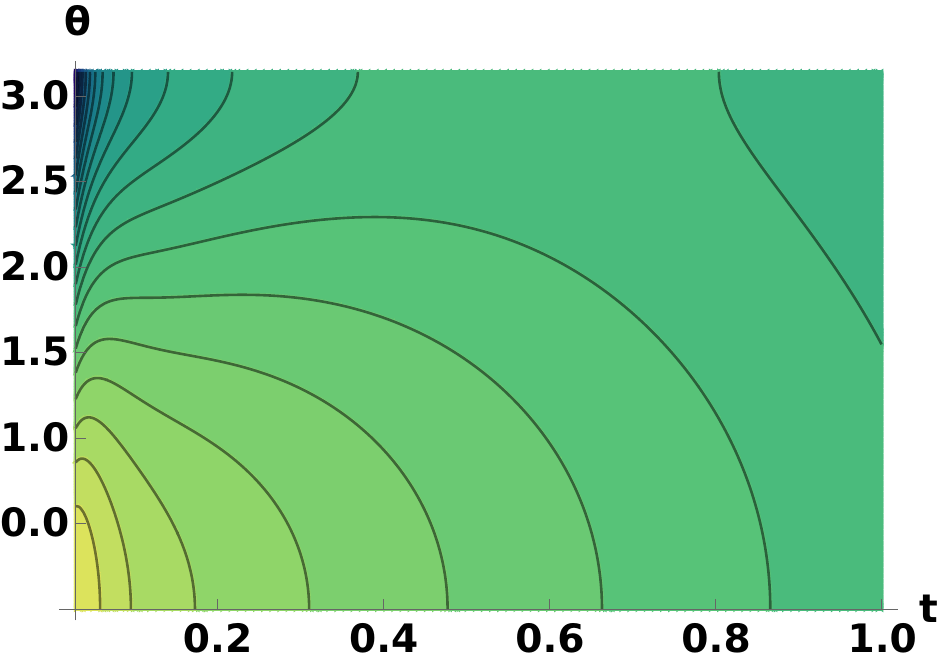}\\
	
\includegraphics[width=60mm]{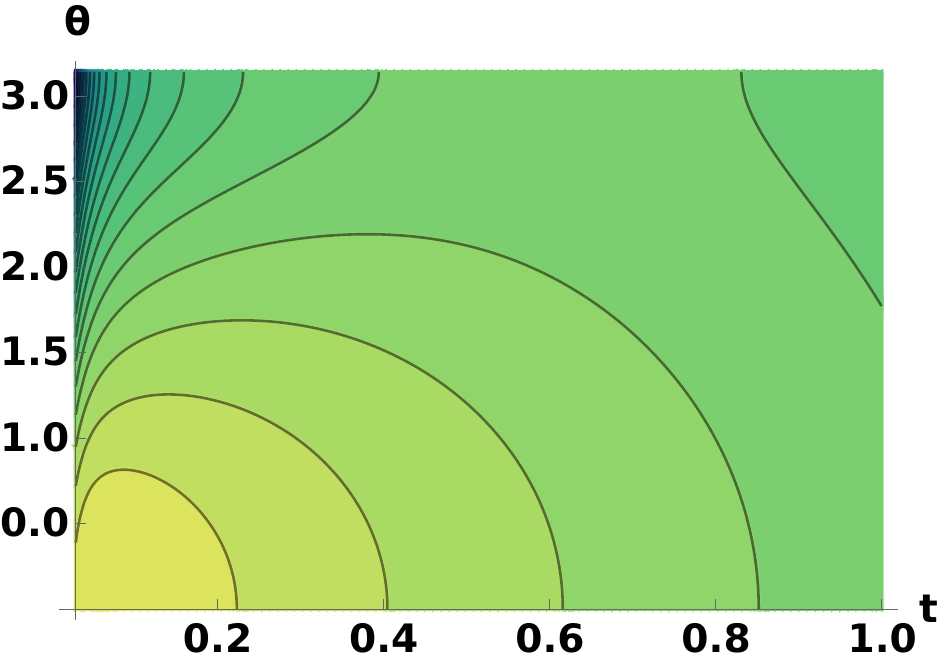}\\

	\caption{Heat maps of the (log of the) joint distribution of the hitting angle $\theta$ and hitting time $t$ (expressed in units of $b^2/D$) for a particle starting its diffusive trajectory at a distance $r_0=1.6,2.6,3.6,4.6$, top to bottom, in an annulus with $a=1,b=6$.}\label{fig:heatlog}
\end{figure}

We present some examples of the joint distribution of the hitting angle and hitting time for a system with $a=1,b=6$ for different values of $r_0$ in Fig. \ref{fig:heatlog}. The contour lines help to clarify the role of hitting angle in the set up of diffusive echo for small starting positions, and its gradual disappearance for higher $r_0$ values.

\begin{figure}[htp]
	
	\centering
	
	\includegraphics[width=70mm]{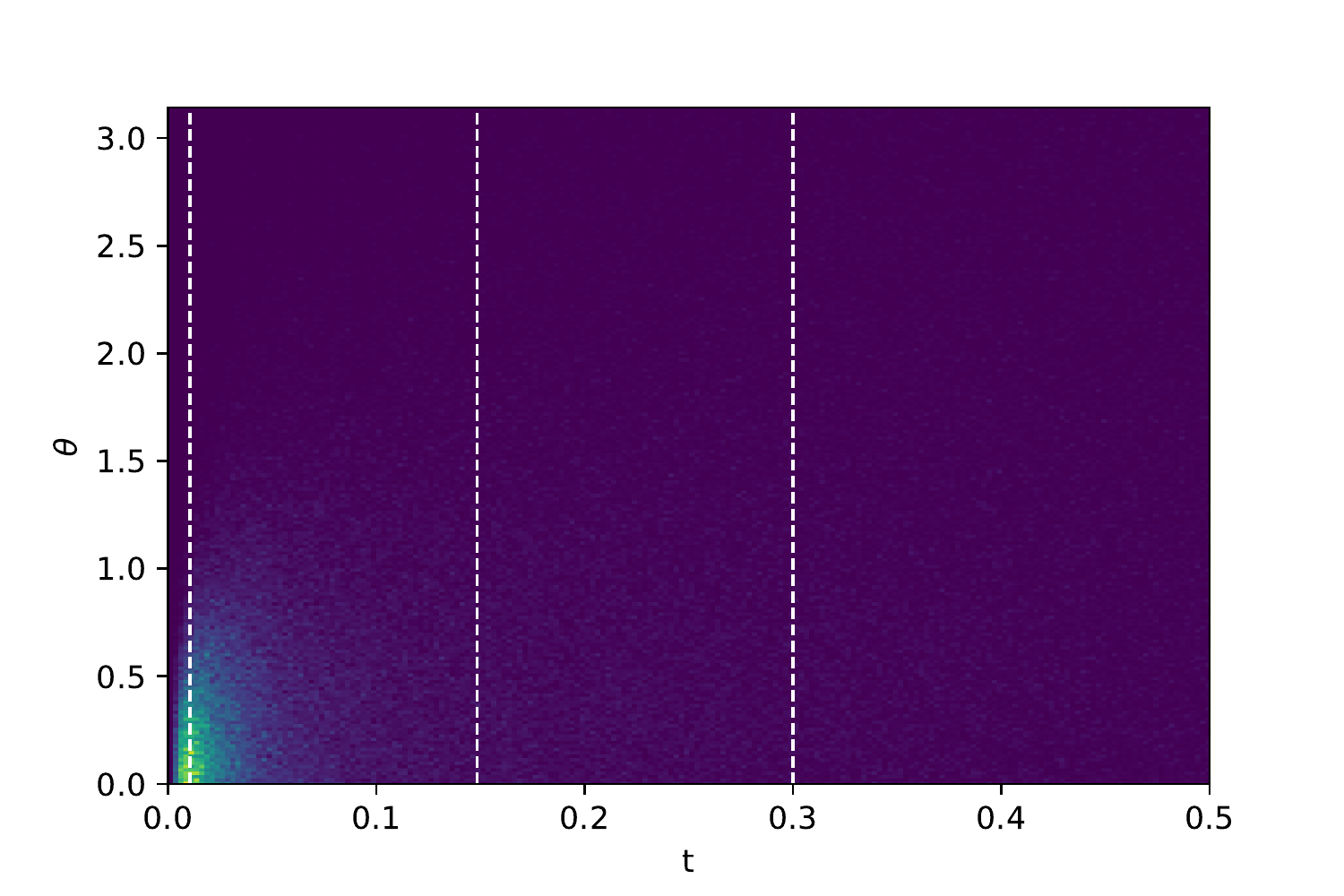}\\
    
	\includegraphics[width=70mm]{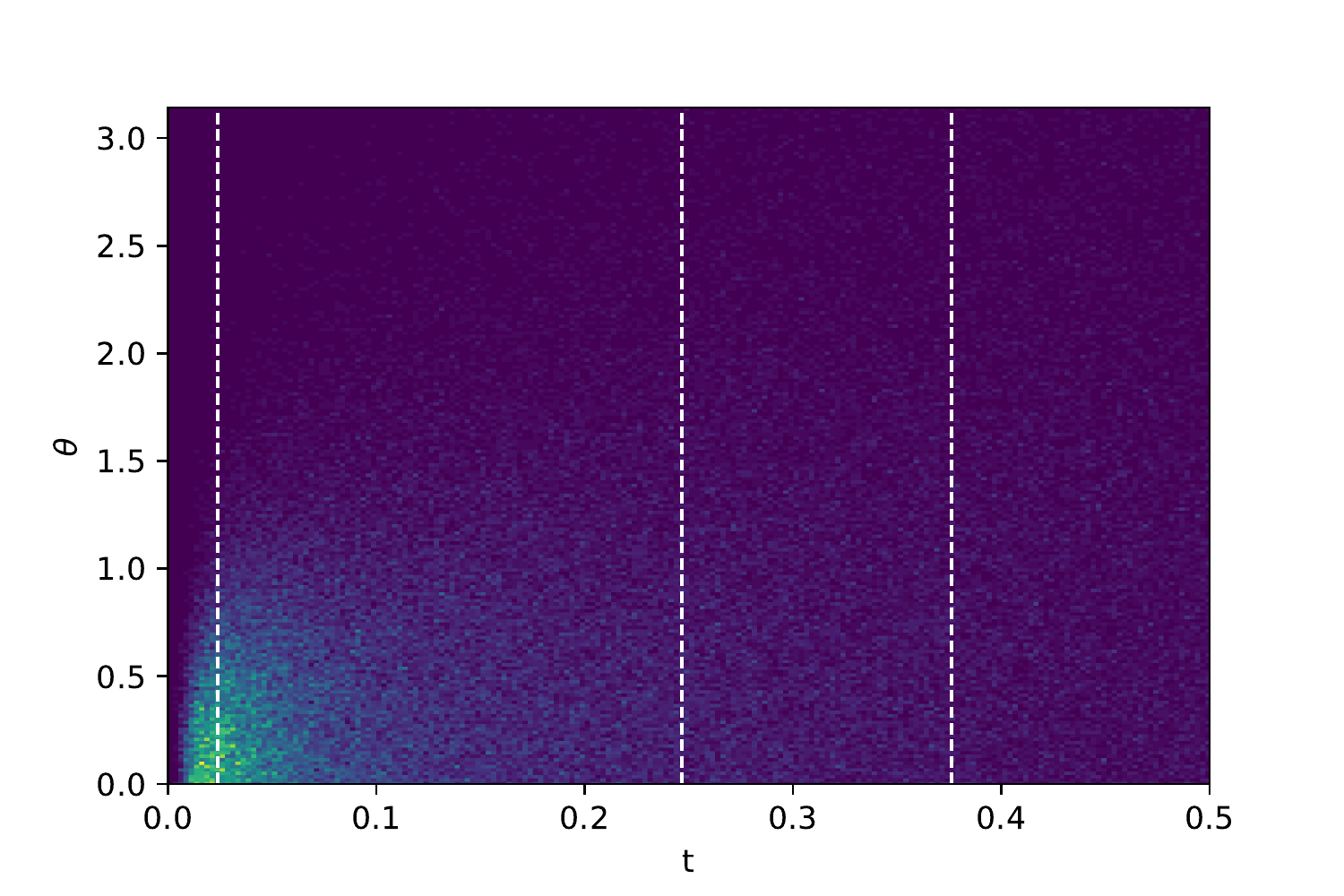}\\

	\includegraphics[width=70mm]{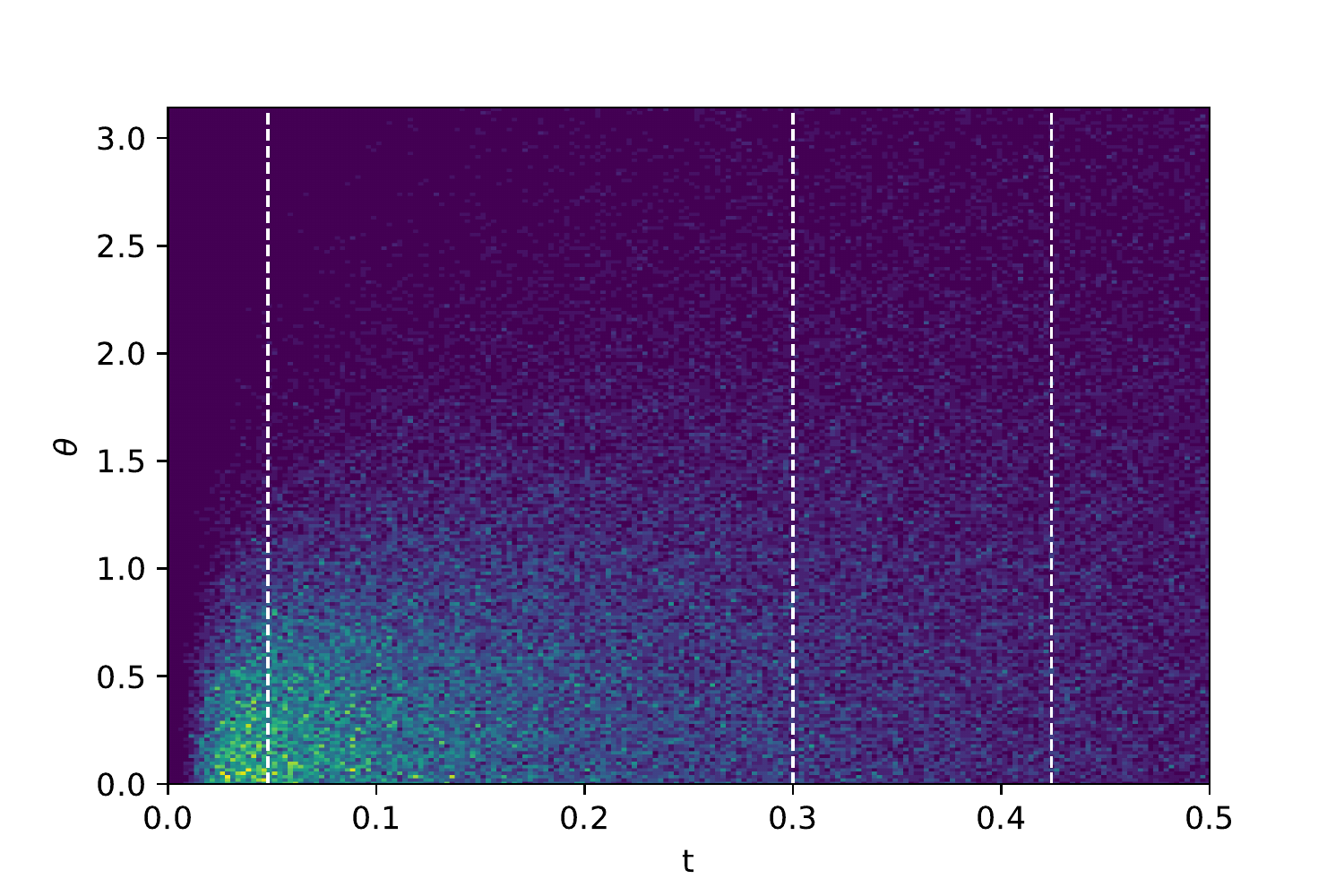}\\

	\includegraphics[width=70mm]{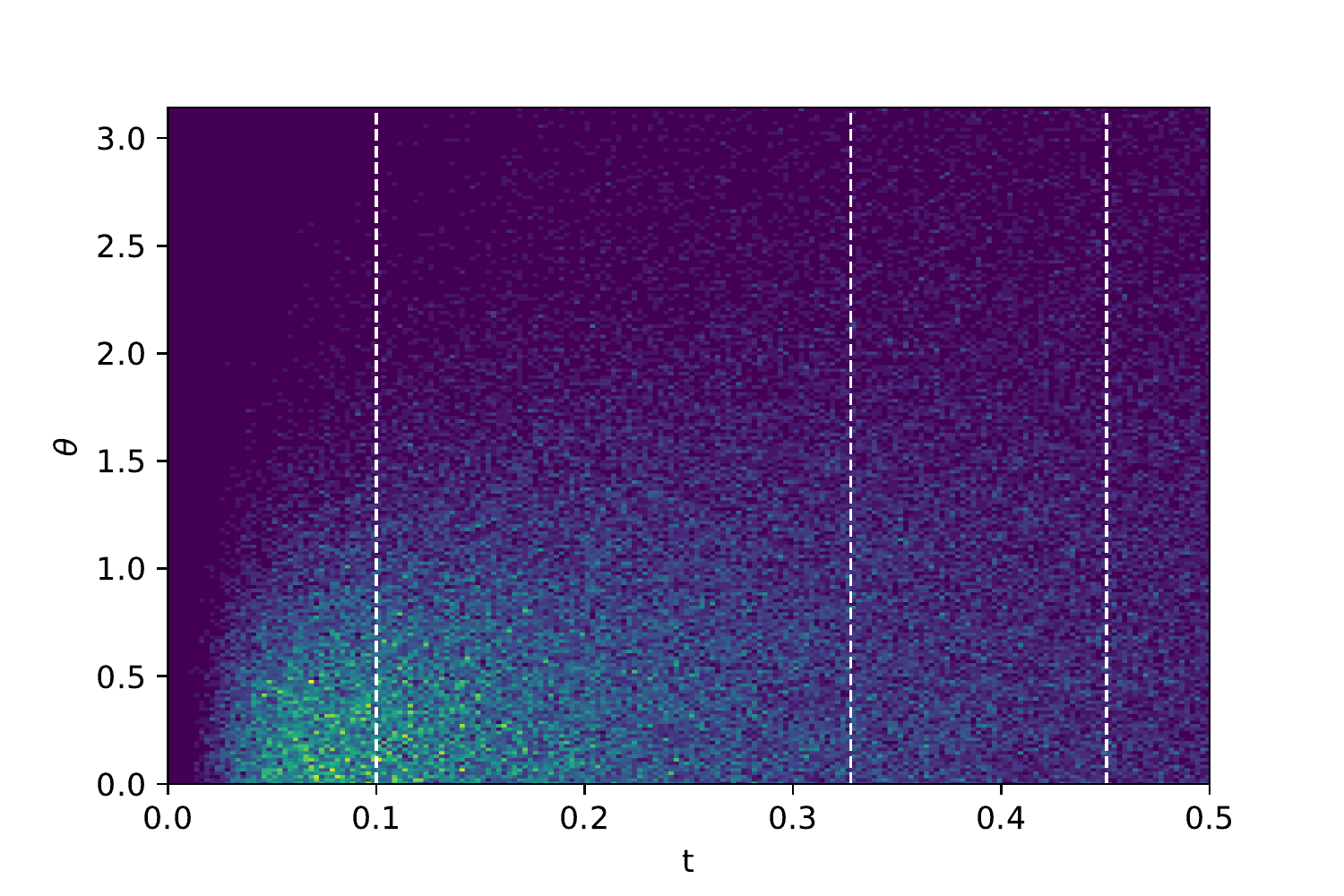}

	\caption{Heat maps of the joint distribution of the hitting angle $\theta$ and hitting time $t$ (expressed in units of $b^2/D$) for a particle starting its diffusive trajectory at a distance $r_0$. The geometry is defined by $a=1,b=4$ and $r_0=2,2.5,3,3.5$, top to bottom. The dashed vertical lines show the most probable, median and mean first passage times from left to right, respectively.}\label{fig:heat}
\end{figure}

For completeness, Fig. \ref{fig:heat} shows heat maps obtained from brownian dynamics simulations for a system with $a=1,b=4$ for various values of $r_0$. Each point shows the hitting angle and hitting time of a single trajectory. These results are also compared with the theoretical values of three typical first passage times to the whole inner absorbing sphere: the most probable $t_{prob}$, mean $\langle t\rangle$ and median $t_{median}$ first passage times (see the Appendix~\ref{app:two}).

\section{Diffusive echo in 3D}

For unbounded diffusive search, it is well known that the dimensionality $d$ of the system plays a major role. The diffusive exploration is compact for $d\leq2$ whereas it is only transient for $d>2$ (Pólya's recurrence theorem \cite{Polya1921,redner2001guide}). For $d=3$, for example, the fraction of particles that never reaches the inner absorbing sphere at $r=a$ is equal to $1-a/{r_0}$. However, if a reflecting barrier is present, these particles are not lost and eventually contribute to the flux on $r=a$, leading to non-intuitive results when the barrier is shifted to infinity \cite{antoine2020nonintuitive}.

The issue of the presence of a diffusive echo in $d>2$ systems is also of interest since one expects seeing two competing effects at stake: the reflection from the external spherical barrier which focuses the diffusive flux toward the target, and the dispersion of the particles in a volume that becomes larger and larger as $d$ increases.

For simplicity we consider the model shown in Fig. \ref{fig:system} where a spherical absorbing surface of radius $a$ is contained within a spherically symmetric reflecting wall of radius $b$. As for the 2D case, our aim is to compute the joint probability that a diffusing particle hits a specific position on the perfectly absorbing central surface.

\begin{figure}[htp]
	
	\centering
	
	\includegraphics[width=85mm]{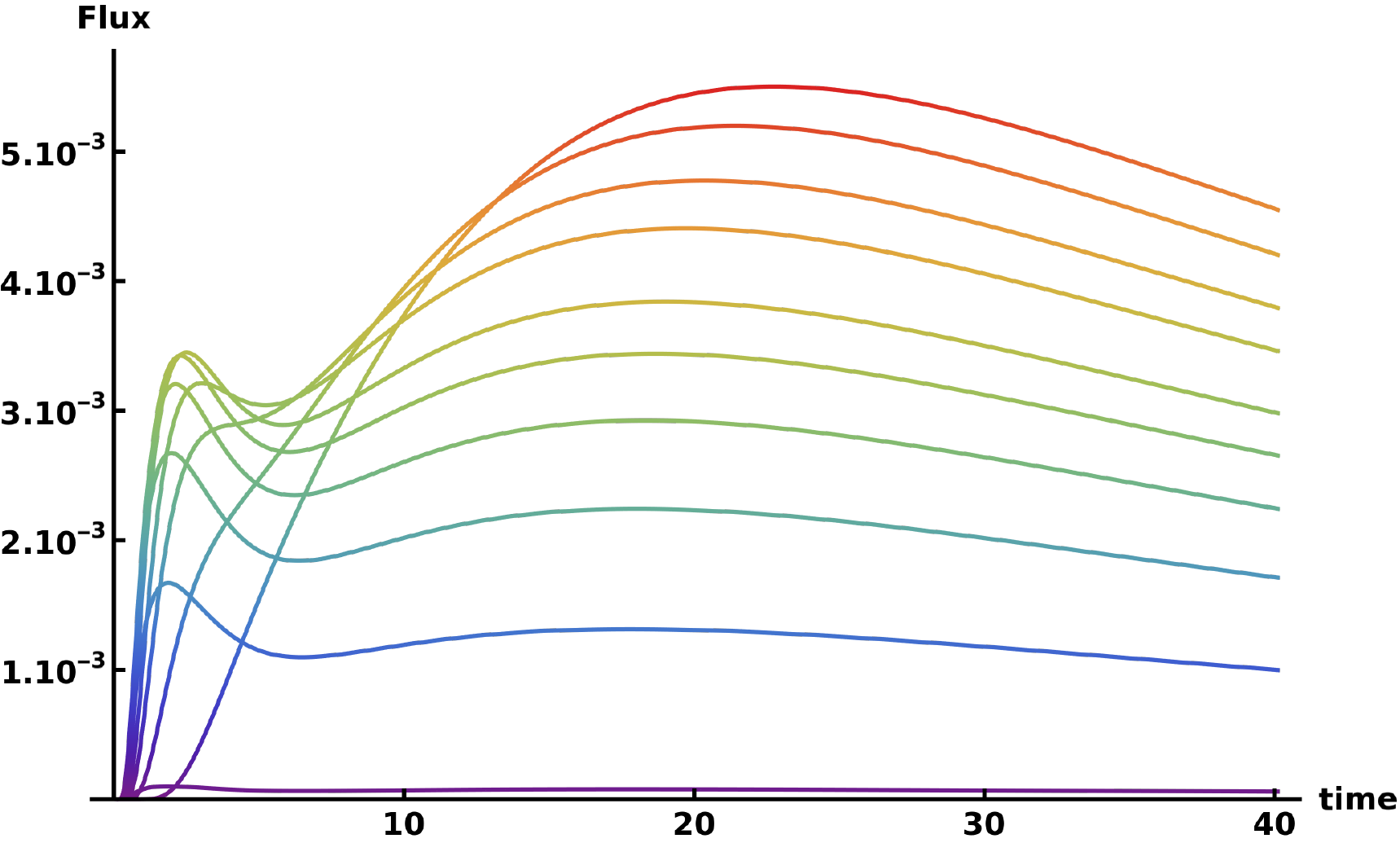}\\
    
	\caption{3D flux $j_{3D}(\theta,t)$ at a hitting angle $\theta=\frac{5 \pi}{6} \simeq 2.6$ for various starting points. From top to bottom $r_0=$ $6$, $3.6$, $2.8$, $2.4$, $2$, $1.8$, $1.6$, $1.4$, $1.2$, $1.01$. The confinement is $b=6$.}\label{fig:CurveFlux3Dvariousr0}
\end{figure}

Due to the spherical symmetry of the system, only one angle $\theta$ is sufficient to label the targeted position with respect to the starting position, and the diffusion equation can be reduced to a simpler equation, similar to the one in 2D. 

As a result, the expression of the flux in more than 2 dimensions is similar to the expression Eq. (\ref{eq:flux2Dmain}) of the 2D flux. For example, in 3D, the main differences are that the $\cos \left( \theta \right)$ term is replaced by a Legendre polynomial in $\cos\theta$ and the Bessel functions are replaced by spherical Bessel functions. The explicit expression for the flux is
\begin{align}\label{eq:flux3Dmain}
j_{3D}\left(\theta,t\right) 
 & =2 a D\sum_{m=0}^{+\infty}\left(m+\frac{1}{2}\right)P_{m}\left(\textrm{cos} \; \theta \right) \nonumber \\
 & \times \sum_{n=1}^{+\infty} \frac{k_{mn}^{3} f_{mn}\left(r_{0}\right)\textrm{e}^{-Dk_{mn}^{2}t}}{c_{mn}}
\end{align}
(see the Appendix~\ref{app:three} for more details).

\begin{figure}[htp]
	\begin{center}
		\resizebox{8.5cm}{!}{\includegraphics{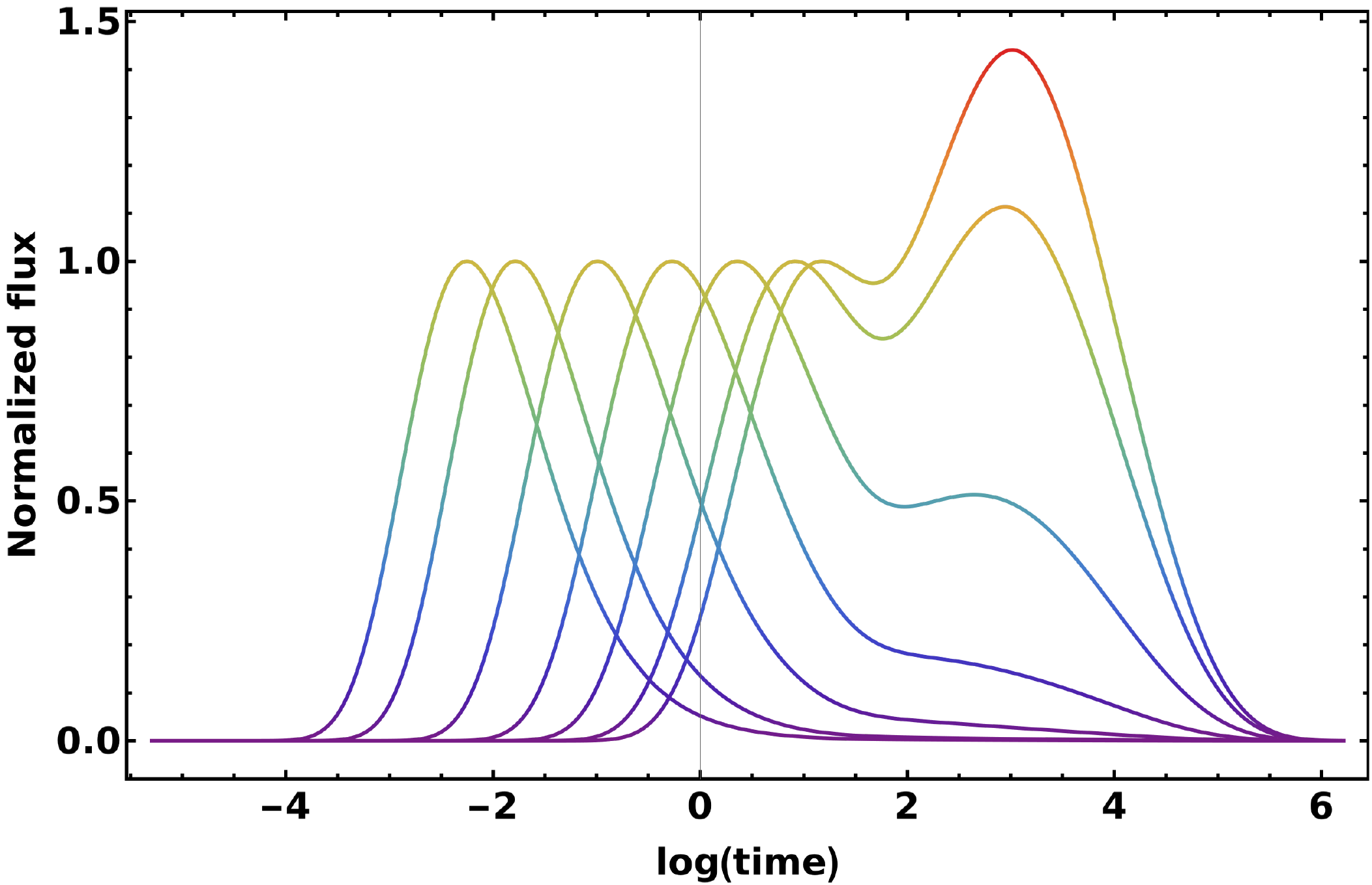}}
	\end{center}
	\caption{3D normalized flux $j_{3D}\left( \theta,t \right) / j_\textrm{first max} \left( \theta,t \right)$ at different hitting angles $\theta$ from the same starting point $r_0 = 1.6$. From left to right $\theta =$ $0$, $\frac{\pi}{6}$, $\frac{\pi}{3}$, $\frac{\pi}{2}$, $\frac{2\pi}{3}$, $\frac{5 \pi}{6}$, $\pi$. The confinement is $b=6$.}\label{fig:CurveFlux3Dvarioustheta}
\end{figure}

The diffusive echo is thus also expected for $d>2$. Fig. (\ref{fig:CurveFlux3Dvariousr0}) and (\ref{fig:CurveFlux3Dvarioustheta}) show how the instantaneous local target flux can present a clear double maximum for a range of starting position $r_0$, targeted position angle $\theta$ and reflecting wall radius $b$.

The diffusive echo is even more pronounced than in 2D, with a more profound dip between the two maxima. The more intense dispersion (in a larger volume) is offset by increased focusing effect of the external spherical barrier.

\section{One dimension and discussion}

The diffusive echo phenomenon is almost imperceptible for one-dimensional ($d=1$) systems \cite{koplik1994tracer,koplik1995universal}. As recalled in the Appendix~\ref{app:one}, for a diffusive process on a segment with reflecting ($r=b$) and  absorbing ($r=a$) boundaries, the expression of the instantaneous flux to the absorbing point is given by
\begin{equation}\label{eq:flux1Dmain}
j_{1D}\left(t\right)  = \frac{2D}{b-a}\sum_{n=1}^{\infty} \textrm{sin}\left[k_{0n} \left( r_{0}-a \right) \right] \, k_{0n} \, \textrm{e}^{-D k_{0n}^{2}t}
\end{equation}
with $k_{0n} = \pi(n-1/2)/(b-a)$.

\begin{figure}[t]
	
	\centering
	
	\includegraphics[width=85mm]{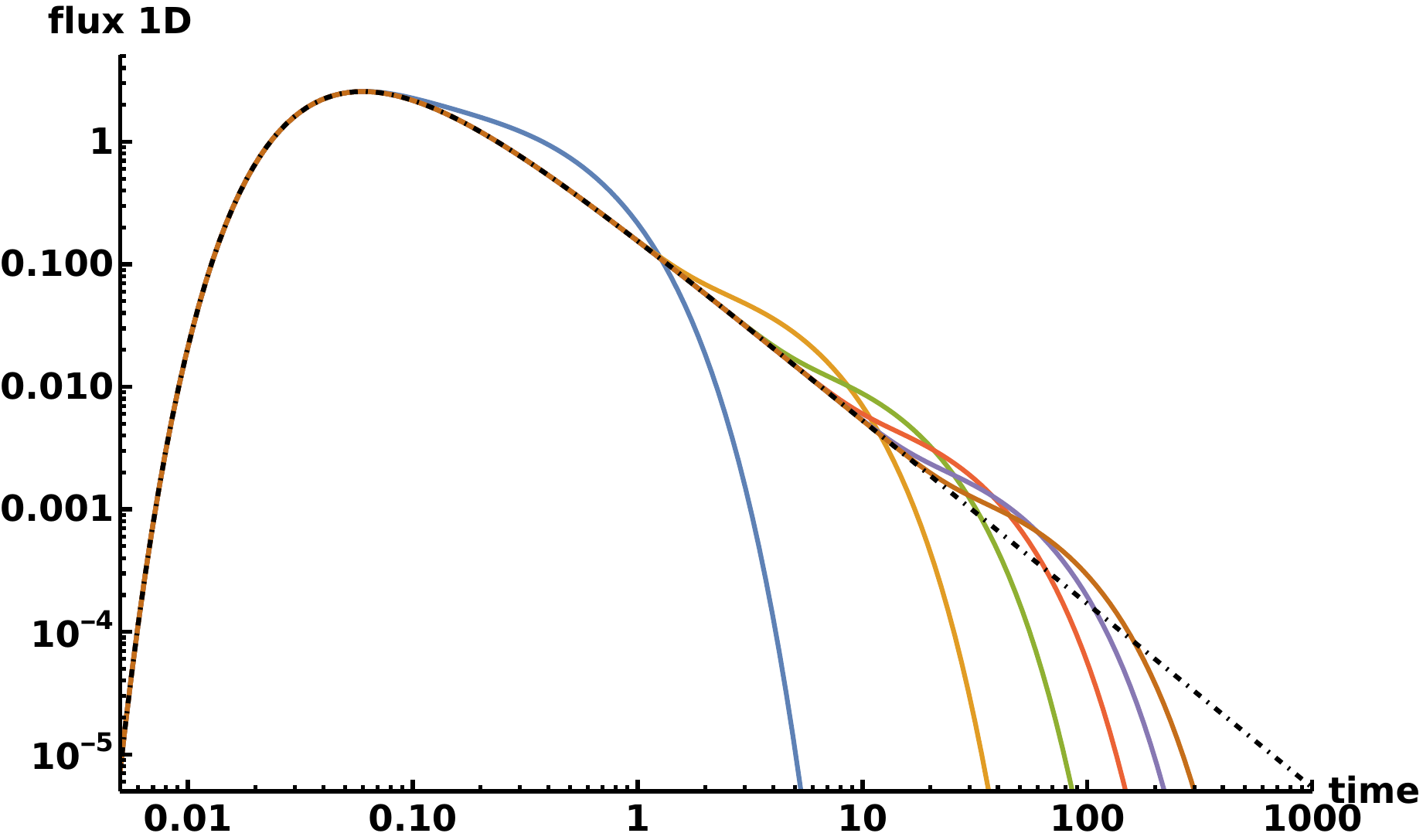}\\

	\caption{1D time dependent flux on the target, for $a=1$ and various values of the confining parameter, $b=2,4,6,8,10,12$, from left to right. The dot-dashed line is for the unbounded case $b=\infty$. The starting point is $r_0 = 1.6$ for all curves.}\label{fig:globalflux1D}
\end{figure} 

Initially, the flux is null, $j_{1D}\left( 0 \right) = 0$, whatever the starting position. It then increases, reaches a maximum and decays exponentially to zero. The exponential time decrease is controlled by the square of the first root, $k_{01}^2$, which is nine times smaller than the square of the other roots. The first term in (\ref{eq:flux1Dmain}) is thus dominant at mid and long term, and no other term can overcome it to create a true echo. The presence of the reflecting boundary is visible only at very long time, in the tail of the flux, where a shoulder is scarcely observable on a log-log scale (see Fig. \ref{fig:globalflux1D} and \cite{koplik1994tracer,koplik1995universal}).

\begin{figure}[h]
	
	\centering
	
	\includegraphics[width=85mm]{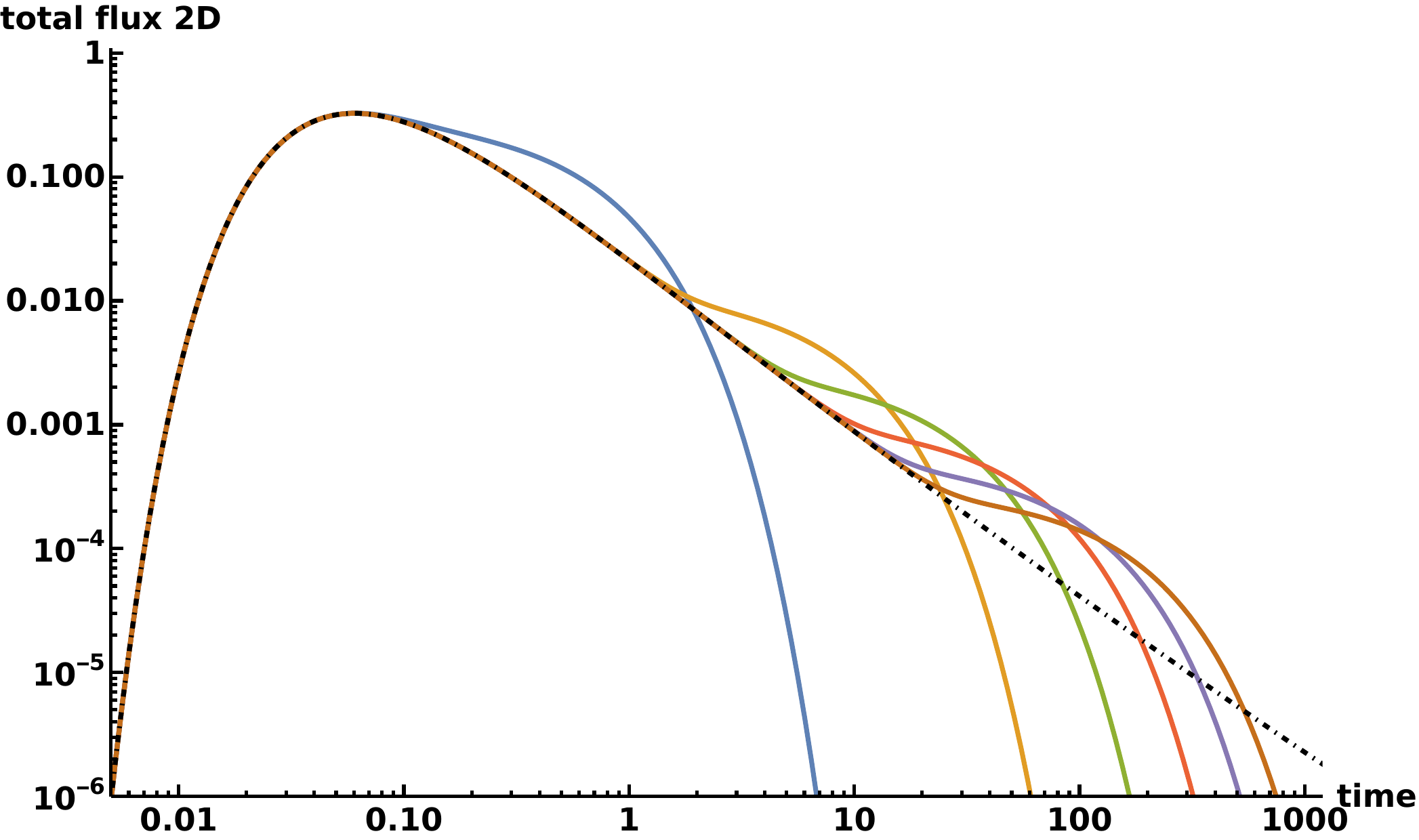}\\

	\caption{Same as Fig. \ref{fig:globalflux1D} except 2D}\label{fig:globalflux2D}
\end{figure} 

\begin{figure}[b]
	
	\centering
	
	\includegraphics[width=85mm]{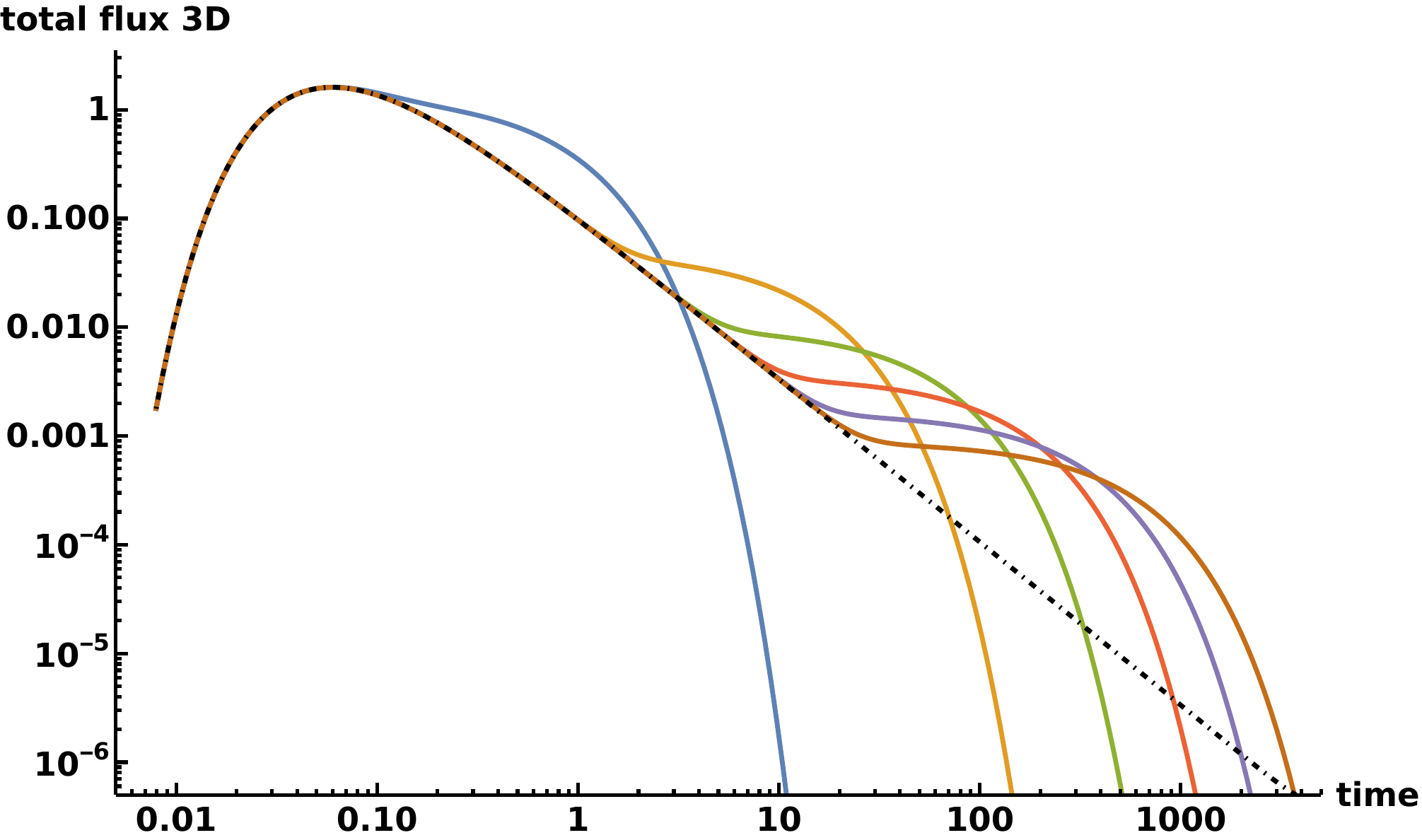}\\

	\caption{Same as Fig. \ref{fig:globalflux1D} except 3D}\label{fig:globalflux3D}
\end{figure}

For an echo to appear,  the presence of an additional variable, like the angle $\theta$ in the 2D and 3D cases described above, appears to be crucial. Depending on its value, this variable may lead to negative terms in the expression of the flux (for example when $\theta \in [\pi/2,\pi]$ for the terms labelled by $m=1$). Similarly to what happens for destructive and constructive interference, one observes a modulation of the flux in the course of time for certain angles.

As one might expect, integrating the flux over the angle $\theta$ (see the Appendixes \ref{app:two} and \ref{app:three} for the resulting expressions) eliminates the echo and one is left with a shoulder on a log-log plot of the flux versus time, as seen in Fig. \ref{fig:globalflux2D} and Fig. \ref{fig:globalflux3D}.

\begin{figure}[htp]
	
	\centering
	
	\includegraphics[width=80mm]{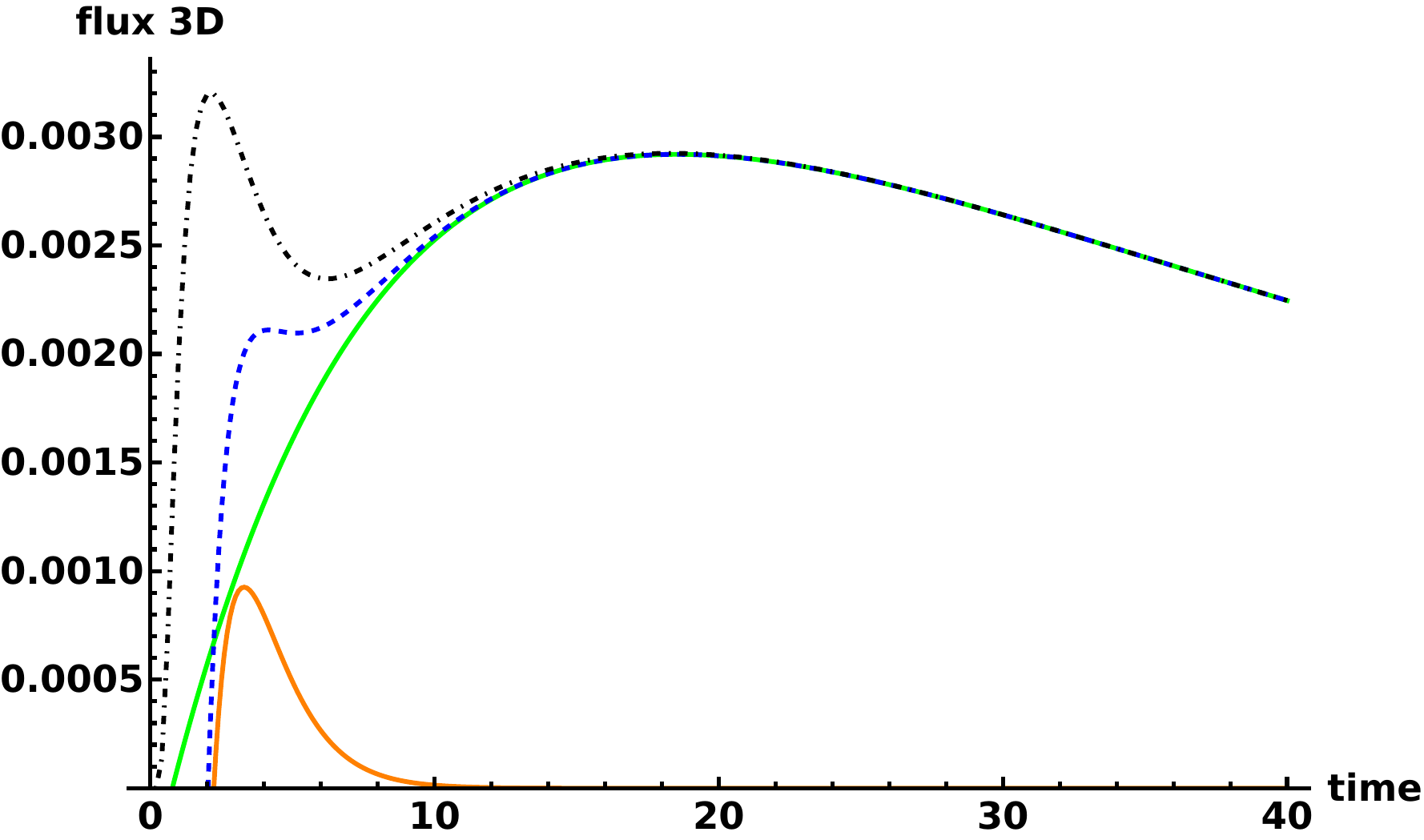}\\

		\caption{The flux $j_{3D}(\theta,t)$ on the spherical absorber at a hitting angle $\theta=5\pi/6$. The starting point is $r_0=1.6$ and the reflecting wall is at $b=6$. The two continuous lines correspond to the partial sums $n=1;m=0,1$ ('indirect')  and $n=2;m=0,1$ ('direct'). The dashed line corresponds to their sum, while the dot-dashed line is the full expression of the flux.}\label{fig:flux3Dfirstterms}
\end{figure}

An examination of the expressions of the fluxes in 2D, Eq. (\ref{eq:flux2Dmain}), and 3D, Eq. (\ref{eq:flux3Dmain}), shows that only a few terms in the sums are needed to account for the echo effect. In 3D for example it suffices to consider the first four terms $n=1,2$ and $m=0,1$ in Eq. (\ref{eq:flux3Dmain}) to get the double maximum in the flux (see Fig. \ref{fig:flux3Dfirstterms}). With the first sixteen terms (i.e. $n$ from $1$ to $4$, and $m$ from $0$ to $3$), one obtains a curve which is almost indistinguishable from the full curve obtained for $n \rightarrow \infty$  and $m \rightarrow \infty$.

The roots $k_{mn}$ that intervene in the flux expressions Eq. (\ref{eq:flux2Dmain}) and Eq. (\ref{eq:flux3Dmain}) are approximately independent of $m$:
\begin{equation}
k_{mn} \simeq \frac{\pi \,  (n-\frac{1}{2})}{b-a},\;\;\forall m
\end{equation}

As a consequence, for each value of $n$, all the terms $\textrm{e}^{-Dk_{mn}^{2}t}$ evolve in a similar way and can be regrouped. The factors $\cos (m\theta)$ in 2D and $P_m \left( \cos \theta \right)$ in 3D may be negative depending on the value of $\theta$, creating the conditions for a bell-shape curve for each packet of terms labelled by the same value of $n$. The first packet ($n=1$) corresponds to the flux reflected from the external confining boundary ( 'indirect'), whereas the second one corresponds to the 'direct' diffusive process towards the target. See Fig. \ref{fig:flux3Dfirstterms}. The intensity of these different contributions to the flux is governed by $r_0$ and $b$.

In summary, and contrary to the acoustic case, a 'true' diffusive echo (i.e. with two well identified maxima in the time-dependent local flux) cannot exist for one dimensional diffusive processes. Rather counter-intuitively, however, the diffusive echo phenomenon intensifies as the dimension increases. Due to the echo, the interval of peak flux can be largely extended in time, showing the relevance of the phenomenon for enhancing the search efficiency in a confined environment.

Although quite simple, the geometry considered here allows one to grasp the essence of the phenomenon. Further research could explore other, more constrained geometries for which multiple diffusive echoes might appear. In particular, the optimum geometry for observing the strongest echo is an open question (as well as the maximum number of observable echoes). The diffusive echo induced by moving obstacles is also worth considering, given its relevance for real diffusive processes. In a similar vein, an effort could be made to determine the role of the absorbing central area in the manifestation of the diffusive echo.

These results could shed light on biological diffusive processes such as those that govern the (passive) search for a specific target in a living, confined environment where traps or lethal zones may impede the search.

{\bf Acknowledgements.} We thank Denis Grebenkov for informing us about his recent work that presents related expressions as well as other references.

\appendix

\section*{Appendixes}

Here we detail the solution of the time-dependent diffusion equations in 1D, 2D and 3D that are necessary to obtain the time dependent hitting angle distribution.

\section{\label{app:one}One dimension}

The (Green's function of the) concentration $c\left(r,t\right)$, $a\leq r\leq b$, obeys the
following diffusion equation 
\begin{equation}
\frac{\partial c}{\partial t'}-\frac{\partial^{2}c}{\partial r'^{2}}=\delta\left(r'-r'_{0}\right)\delta\left(t'\right)
\end{equation}
with the boundary condition $c\left(a,t\right)=0$, $\partial c/\partial r\mid_{r=b}=0$
in terms of dimensionless variables $r'=r/a$ and $t'=Dt/a^{2}$. In the following we drop the primes for simplicity.

This  one dimensional problem can be solved simply, but here we use a slightly more complicated method  that parallels the treatment of the other dimensions.

Introducing the Laplace transform $\widehat{f}\left(s\right)=\int_{0}^{\infty}dt\,\textrm{e}^{-st}c\left(r,t\right)$,
one obtains the following differential equation
\begin{equation}
\frac{d^{2}\widehat{f}}{dx^{2}}-\widehat{f}=-\frac{\delta\left(x-x_{0}\right)}{\sqrt{s}}
\end{equation}
where we have introduced the reduced variable $x=r\sqrt{s}$.

The boundary conditions $\widehat{f}\left(x_{a}\right)=0$ and $\frac{d\widehat{f}}{dx}\left(x_{b}\right)=0$
imply that the global solution can be written as the combination of
interior and exterior solutions 
\begin{equation}
\begin{cases}
\widehat{f}_{<}\left(x\right) & =A \, \textrm{sinh}\left(x-x_{a}\right),  \; x<x_{0}\\
\widehat{f}_{>}\left(x\right) & =B \, \textrm{cosh}\left(x_{b}-x\right), \; x>x_{0}
\end{cases}
\end{equation}

Then, by continuity of $\widehat{f}$ and integration of (2) across
the discontinuity of $d\widehat{f}/dx$ in $x_{0}$, one finally gets
\begin{equation}
\widehat{f}\left(x\right)=\frac{\textrm{sinh}\left(x_{<}-x_{a}\right)\textrm{cosh}\left(x_{b}-x_{>}\right)}{\sqrt{s} \; \textrm{cosh}\left(x_{b}-x_{a}\right)}
\end{equation}
with $x_{<}=\textrm{min}\left(x,x_{0}\right)$ and $x_{>}=\textrm{max}\left(x,x_{0}\right)$.\\

The inverse Laplace transform and residue theorem provide the global
solution
\begin{align}
c\left(r,\theta,t\right) & =\frac{1}{2\pi i}\int_{\gamma-i\infty}^{\gamma+i\infty}ds\,\textrm{e}^{Dst}\,\widehat{f}\left(x\right)\nonumber \\
 & =\sum_{n}\textrm{Res}\left[\textrm{e}^{Dst}\,\widehat{f}\left(x\right)\right]
\end{align}
where the sum of the residues is over the singularities of $\widehat{f}$,
i.e. over the roots $s_{n}$ of the auxiliary equation 
\begin{equation}
g\left(s\right)=\textrm{cosh}\left(\sqrt{s}\left(b-a\right)\right)=0
\end{equation}
which are 
\begin{equation}
\sqrt{s_{n}}=-i k_{0n},\;n\in\mathbb{N}^*
\end{equation}
with 
\begin{equation}
k_{0n} = \frac{\pi \,  (n-\frac{1}{2})}{b-a}
\end{equation}

The term $g'\left(s_{n}\right)$ needed to calculate the residue is 
\begin{equation}
g'\left(s_{n}\right)=\frac{\left(b-a\right) }{2 k_{0n}} \left(-1\right)^{n+1}
\end{equation}
and one finally obtains 
\begin{equation}
c\left(r,t\right)  =\frac{2}{b-a}\sum_{n=1}^{\infty}\textrm{e}^{-D k_{0n}^2 t} \, \textrm{sin}\left[ k_{0n} \left( r-a \right) \right]\textrm{sin}\left[k_{0n} \left( r_{0}-a \right) \right]
\end{equation}
where the completeness relation of the basis functions $\textrm{sin}\left[k_{0n} \left( r_{0}-a \right) \right]$ reads
\begin{align}
c\left(r,0\right) & =\frac{2}{b-a}\sum_{n=1}^{\infty} \textrm{sin}\left[ k_{0n} \left( r-a \right) \right]\textrm{sin}\left[k_{0n} \left( r_{0}-a \right) \right] \nonumber \\
& = \delta\left(r-r_{0}\right)
\end{align}

The instantaneous flux in $r=a$ is 
\begin{align}
j_{1D}\left(t\right) & = \left. D\frac{\partial c}{\partial r} \right| _{a} \nonumber \\
 & =\frac{2D}{b-a}\sum_{n=1}^{\infty} \textrm{sin}\left[k_{0n} \left( r_{0}-a \right) \right] \, k_{0n} \, \textrm{e}^{-D k_{0n}^{2}t}
\end{align}
and the cumulative hitting probability $J_{1D}\left(t\right)  =\int_{0}^{t}j_{1D}\left(t'\right)dt'$ at time $t$ is 
\begin{equation}
J_{1D}\left(t\right) =\frac{2}{b-a} \sum_{n=1}^{\infty}\left(1-\textrm{e}^{-D k_{0n}^{2} t}\right) \frac{\textrm{sin}\left[k_{0n} \left( r_{0}-a \right) \right]}{k_{0n}}
\end{equation}
which converges monotonically to $1$ when $t\longrightarrow\infty$.

In the limit of an infinitely distant confining wall, $b\longrightarrow\infty$,
we recover the usual unbounded expressions
\begin{equation}
c^{*}\left(r,t\right)=\frac{1}{2\sqrt{\pi Dt}}\left(\textrm{e}^{-\frac{\left(r-r_{0}\right)^{2}}{4Dt}}-\textrm{e}^{-\frac{\left(2a-r-r_{0}\right)^{2}}{4Dt}}\right)
\end{equation}
and
\begin{equation}
J_{1D}^{*}\left(t\right)=\int_{0}^{t}j^{*}\left(t'\right)dt'=1-\textrm{erf}\left(\frac{r_{0}-a}{2\sqrt{Dt}}\right)
\end{equation}

\section{\label{app:two}Two dimensions}

The concentration $c\left(r,\theta,t\right)$, $a\le r\le b$, $-\pi\le \theta \le \pi$, where $r$ is the radial distance from the center of the target and $\theta$ is the azimuthal angle, obeys the following diffusion equation 

\begin{equation}\label{eq:diffusion2D}
\frac{\partial c}{\partial t'}-\frac{\partial^{2}c}{\partial r'^{2}}-\frac{1}{r'}\frac{\partial c}{\partial r'}-\frac{1}{r'^{2}}\frac{\partial^{2}c}{\partial\theta^{2}}=\frac{\delta\left(r'-r'_{0}\right)\delta\left(\theta\right)\delta\left(t'\right)}{r'_{0}},
\end{equation}
with the boundary conditions $c(a,\theta,t)=0$, $\partial c/\partial r|_{r=b}=0$
in terms of dimensionless variables $r'=r/a$ and $t'=Dt/a^2$. In the following we drop the primes for simplicity.

Since the angle dependence is symmetric in $\theta$, the
solution is necessarily of the form 
\begin{equation}
c(r,\theta,t)=\sum_{k=-\infty}^{+\infty}\textrm{cos}(k\theta)\,f_{k}(r,t)
\end{equation}
Multiplying both sides of (\ref{eq:diffusion2D}) by $\textrm{cos}(m\theta)$ and integrating over $\theta$ leads to
\begin{equation}
\frac{\partial f_{m}}{\partial t}-\frac{\partial^{2}f_{m}}{\partial r^{2}}-\frac{1}{r}\frac{\partial f_{m}}{\partial r}+\frac{m^{2}}{r^{2}}f_{m}=\frac{\delta\left(r-r_{0}\right)\delta\left(t\right)}{2 \pi r_{0}}
\end{equation}
since $\int_{-\pi}^{+\pi} d\theta \;\textrm{cos}(k\theta)\,\textrm{cos}(m\theta) = \pi \delta_{k,\left| m \right| }$ and $f_{-m} = f_m$.
Introducing the Laplace transform $\widehat{f}_{m}\left(x\right)=\int_{0}^{\infty}dt\,\textrm{e}^{-st}f_{m}\left(x,t\right)$, one finally obtains the following well-known differential equation 
\begin{equation}\label{eq:diffusion2Dfmhat}
\frac{d^{2}\widehat{f}_{m}}{dx^{2}}+\frac{1}{x}\frac{d\widehat{f}_{m}}{dx}-\left(1+\frac{m^{2}}{x^{2}}\right)\widehat{f}_{m}=-\frac{\delta\left(x-x_{0}\right)}{2\pi x_{0}}
\end{equation}
where the reduced variable $x=r\sqrt{s}$ has been introduced.

The boundary conditions 
\begin{equation}
\widehat{f}_{m}\left(x_{a}\right)=0\qquad {\rm and}\qquad\frac{\partial\widehat{f}_{m}}{\partial x}\left(x_{b}\right)=0,
\end{equation}
corresponding to absorption at $r=a$ and reflection at $r=b$ respectively, imply
that the global solution can be written as the combination of interior
and exterior solutions:
\begin{align}
\widehat{f}_{m<}\left(x\right) & \underset{x<x_{0}}{=}A_{m}\left(I_{m}\left(x\right)K_{m}\left(x_{a}\right)-K_{m}\left(x\right)I_{m}\left(x_{a}\right)\right) \nonumber\\
\widehat{f}_{m>}\left(x\right) & \underset{x>x_{0}}{=}B_{m}\left(I_{m}\left(x\right)K_{m}'\left(x_{b}\right)-K_{m}\left(x\right)I_{m}'\left(x_{b}\right)\right)
\end{align}
in terms of $I_{m}\left(x\right)$ and $K_{m}\left(x\right)$, the
modified Bessel functions of the first and second kind of order $m$.

Then, by continuity of $\widehat{f}_{m}$ and integration of (\ref{eq:diffusion2Dfmhat}) across
the discontinuity in $x_{0}$, one finally gets 
\begin{equation}\label{eq:2Dfmhat}
\widehat{f}_{m}\left(x\right)=\frac{1}{2\pi}\frac{\left(I_{mb}'K_{m>}-K_{mb}'I_{m>}\right)\left(I_{m<}K_{ma}-K_{m<}I_{ma}\right)}{I_{mb}'K_{ma}-K_{mb}'I_{ma}}
\end{equation}
with $x_{<}=\textrm{min}\left(x,x_{0}\right)$ and $x_{>}=\textrm{max}\left(x,x_{0}\right)$,
and where, for example,
$I_{mb}'$ and $K_{m>}$ stand  for $\frac{\partial I_{m}}{\partial x}\left(x_{b}\right)$
and $K_{m}\left(x_{>}\right)$, respectively.

The inverse Laplace transform and residue theorem provide the global
solution in the time domain in terms of dimensional variables
\begin{align}
c\left(r,\theta,t\right) &= \sum_{m=-\infty}^{+\infty}\frac{\textrm{cos}\left(m\left(\theta-\theta_{0}\right)\right)}{2\pi i}\int_{\gamma-i\infty}^{\gamma+i\infty}ds\,\textrm{e}^{Dst}\,\widehat{f}_{m}\left(x\right) \nonumber\\
&= \sum_{m=-\infty}^{+\infty}\textrm{cos}\left(m\left(\theta-\theta_{0}\right)\right)\,\sum_{n}\textrm{Res}\left[\textrm{e}^{Dst}\,\widehat{f}_{m}\left(x\right)\right]
\end{align}
where the sum of the residues is over the singularities of $\widehat{f}_{m}$,
i.e. over the roots $s_{n}$ of the auxiliary equation 
\begin{equation}\label{eq:2Dgs}
g\left(s\right) = I_{m}'\left(qb\right)K_{m}\left(qa\right)-K_{m}'\left(qb\right)I_{m}\left(qa\right) = 0
\end{equation}
where $q=\sqrt{s}$.\\

The term $g'\left(s_{n}\right)$ appearing in the residue is 
\begin{align}
g'\left(s_{n}\right)&=\frac{b}{2q_{n}}\left(I_{mb}''K_{ma}-K_{mb}''I_{ma}\right)\nonumber\\
&+\frac{a}{2q_{n}}\left(I_{mb}'K_{ma}'-K_{mb}'I_{ma}'\right)
\end{align}
(in shorthand notation here, but evaluated in $s=s_{n}$), which can be
simplified thanks to (\ref{eq:2Dgs}) as 
\begin{align}
g'\left(s_{n}\right)&=\frac{b}{2q_{n}}\frac{I_{ma}}{I_{mb}'}\left(I_{mb}''K_{mb}'-K_{mb}''I_{mb}'\right)\nonumber\\
&+\frac{a}{2q_{n}}\frac{I_{mb}'}{I_{ma}}\left(I_{ma}K_{ma}'-K_{ma}I_{ma}'\right)
\end{align}
Then, using the Bessel identities 
\begin{equation}\label{eq:BesselIdentity}
I_{m}\left(q_{n}a\right)K_{m}'\left(q_{n}a\right)-K_{m}\left(q_{n}a\right)I_{m}'\left(q_{n}a\right)=-\frac{1}{q_{n}a}
\end{equation}
\begin{equation}
I_{m}''\left(q_{n}b\right)K_{m}'\left(q_{n}b\right)-K_{m}''\left(q_{n}b\right)I_{m}'\left(q_{n}b\right)=-\frac{\left(q_{n}b\right)^{2} + m^{2}}{\left(q_{n}b\right)^{3}}
\end{equation}
the function $g'\left(s_{n}\right)$ reads 
\begin{equation}
g'\left(s_{n}\right)=-\frac{1}{2q_{n}^{2}}\left(\frac{I_{mb}'}{I_{ma}}+\frac{I_{ma}}{I_{mb}'}\left(1+\left(\frac{m}{q_{n}b}\right)^{2}\right)\right)
\end{equation}

Replacing in (\ref{eq:2Dfmhat}) $K_{mb}'$ by $I_{mb}'K_{ma}/I_{ma}$ thanks to (\ref{eq:2Dgs}),
one obtains 
\begin{align}
c\left(r,\theta,t\right) & =\sum_{m=-\infty}^{+\infty}\frac{\textrm{cos}\left(m\theta\right)}{\pi}\,\sum_{n}\textrm{e}^{Dq_{n}^{2}t}\,q_{n}^{2}\nonumber \\
 & \times \frac{\left(K_{m>}I_{ma}-I_{m>}K_{ma}\right)\left(K_{m<}I_{ma}-I_{m<}K_{ma}\right)}{1+\left(\frac{I_{ma}}{I_{mb}'}\right)^{2}\left(1+\left(\frac{m}{q_{n}b}\right)^{2}\right)}
\end{align}
where the product $\left(K_{m>}I_{ma}-I_{m>}K_{ma}\right)$ $\left(K_{m<}I_{ma}-I_{m<}K_{ma}\right)$
may also be written as $\left(K_{mr}I_{ma}-I_{mr}K_{ma}\right)\left(K_{mr_{0}}I_{ma}-I_{mr_{0}}K_{ma}\right)$
in terms of the variables $r$ and $r_{0}$.\\

Since the solutions $q_{n}=\sqrt{s_{n}}$ of $g\left(s\right)=0$
are purely imaginary: $q_{n}=i\,k_{n}$, with $k_{n}\in\mathbb{R}$,
one can switch to the Bessel functions of the first and second kind,
$J_{m}$ and $Y_{m}$, using the relations 
\begin{align}
I_{m}\left(ik_{n}u\right) & =i^{m}J_{m}\left(k_{n}u\right) \nonumber \\
I_{m}'\left(ik_{n}u\right) & =i^{m-1}J_{m}'\left(k_{n}u\right) \\
K_{m}\left(ik_{n}u\right) & =\frac{\pi}{2}\left(-i\right)^{m}\left[-iJ_{m}\left(k_{n}u\right)-Y_{m}\left(k_{n}u\right)\right]\nonumber
\end{align}
for $m$ integer in the last expression.

The final solution gives back the expression (\ref{eq:nrthetat}):
\begin{align}
c\left(r,\theta,t\right) & =\sum_{m=-\infty}^{+\infty}\frac{\pi}{2}\,\textrm{cos}\left(m\theta\right) \nonumber \\
& \times \sum_{n=1}^{+\infty}\textrm{e}^{-Dk_{mn}^{2}t}\,k_{mn}^{2} \frac{F_{mn}\left(r\right)F_{mn}\left(r_{0}\right)}{C_{mn}}
\end{align}
with the coefficients $C_{mn}$ given by
\begin{equation}
C_{mn} = 2 \left[ \left( \frac{J_{m}(k_{mn} a)}{J_{m}'(k_{mn} b)} \right) ^2 \left( 1 - \left(\frac{m}{k_{mn} b}\right)^2 \right) -  1 \right]
\end{equation}
with
\begin{equation}
F_{mn}\left(r\right)=Y_{m}\left(k_{mn}r\right)J_{m}\left(k_{mn}a\right) - J_{m}\left(k_{mn}r\right)Y_{m}\left(k_{mn}a\right)
\end{equation}
where $k_{mn}$ is the $n^{th}$ solution of the $m^{th}$ auxiliary equation 
\begin{equation}
Y_{m}'\left(kb\right)J_{m}\left(ka\right) - J_{m}'\left(kb\right)Y_{m}\left(ka\right)=0
\end{equation}

The completeness relation of the basis functions $k_{mn} F_{mn}\left(r\right) / \sqrt{C_{mn}}$, reads (whatever the value of $m$)
\begin{align}
\sum_{n=1}^{+\infty}\,{\pi^2} \, k_{mn}^{2} \frac{F_{mn}\left(r\right)F_{mn}\left(r_{0}\right)}{C_{mn}} = \frac{\delta\left(r-r_{0}\right)}{r_0 }
\end{align}
for any value of $m$, since $\sum_{m=-\infty}^{+\infty}\textrm{cos}\left(m\theta\right) = 2 \pi \delta\left(\theta\right)$ in the $c\left(r,\theta,0\right)$ expression.

The instantaneous flux at a specific point $\left(r=a,\theta\right)$
of the absorbing inner boundary is 
\begin{align}
j\left(\theta,t\right) & =\left. aD\frac{\partial c}{\partial r}\right| _{a} \nonumber \\
 & =D\sum_{m=-\infty}^{+\infty}\sum_{n=1}^{+\infty}\frac{\textrm{cos}\left(m\theta\right)F_{mn}\left(r_{0}\right)k_{mn}^{2}\textrm{e}^{-Dk_{mn}^{2}t}}{C_{mn}}
\end{align}
since $\frac{\partial F_{mn}}{\partial r} \left( a \right) = 2/\pi a$ thanks to the Bessel identity (\ref{eq:BesselIdentity}).

The cumulative hitting probability $J\left(\theta,t\right)=\int_{0}^{t}j\left(\theta,t'\right)dt'$ in $\theta$ at time $t$ is 
\begin{equation}
J\left(\theta,t\right)=\sum_{m=-\infty}^{+\infty}\sum_{n=1}^{+\infty}\frac{\textrm{cos}\left(m\theta\right)F_{mn}\left(r_{0}\right)\left(1-\textrm{e}^{-Dk_{mn}^{2}t}\right)}{C_{mn}}
\end{equation}
whose integral over $\theta$
\begin{equation}
\int_{-\pi}^{\pi}J\left(\theta,t\right)d\theta=2\pi\sum_{n=1}^{+\infty}\frac{F_{0n}\left(r_{0}\right)\left(1-\textrm{e}^{-Dk_{0n}^{2}t}\right)}{C_{0n}}
\end{equation}
converges monotonically to $1$ for $t\longrightarrow\infty$ (whatever the value of $r_0$).

The total flux to the inner absorbing circle can also be computed:
\begin{equation}
j_{tot}\left(t\right) = \int_{-\pi}^{\pi}j\left(\theta,t\right)d\theta=2\pi D \sum_{n=1}^{+\infty}\frac{F_{0n}\left(r_{0}\right) k_{0n}^{2} \textrm{e}^{-Dk_{0n}^{2}t}}{C_{0n}}
\end{equation}
as well as the total survival probability
\begin{equation}
S_{tot}\left(t\right) = \int_{-\pi}^{\pi} \int_{a}^{b} c\left(r,\theta,t\right) r dr d\theta=2\pi \sum_{n=1}^{+\infty}\frac{F_{0n}\left(r_{0}\right) \textrm{e}^{-Dk_{0n}^{2}t}}{C_{0n}}
\end{equation}
from which can be inferred the typical first passage times of the process: the mean first passage time
\begin{equation}
< t > = \int_{0}^{\infty} S_{tot}\left(t\right)  dt = \frac{2\pi}{D} \sum_{n=1}^{+\infty}\frac{F_{0n}\left(r_{0}\right)} {k_{0n}^{2} C_{0n}}
\end{equation}
the median first passage time, $t_{median}$, satisfying $S_{tot}\left(t_{median}\right) = 1/2$, and the most probable first passage time, $t_{prob}$, for which $\frac{\partial j_{tot} }{\partial t} \left(t_{prob} \right) = 0$.

The eventual hitting probability distribution, $\tilde{J}\left(\theta\right)$ $=\lim_{t \rightarrow + \infty} \left( J\left(\theta,t\right) \right) $,
can be obtained directly from the Laplace transform of $c\left(r,\theta,t\right)$:
\begin{align}
&\tilde{J}\left(\theta\right)  =\left. a D\int_{0}^{\infty}dt\;\frac{\partial c}{\partial r}\right|_{r=a}\nonumber \\
& =aD\frac{\partial}{\partial r}\widehat{c}\left(r,\theta,s\right)\mid_{r=a,s\rightarrow0}\nonumber \\
&=\left.\frac{1}{2\pi}\sum_{m=-\infty}^{+\infty}\textrm{cos}\left(m\theta\right)\,\left(\frac{I_{m0}K_{mb}'-K_{m0}I_{mb}'}{I_{ma}K_{mb}'-K_{ma}I_{mb}'}\right)\right|_{s\rightarrow0}
\end{align}
which gives Eq. (\ref{eq:bound2D}).

In the limit of an infinitely distant confining wall, $b\longrightarrow\infty$,
we recover the usual unbounded expressions
\begin{align}
c^{*}\left(r,\theta,t\right) & =\sum_{m=-\infty}^{+\infty}\frac{\textrm{cos}\left(m\theta\right)}{2\pi} \nonumber \\
 & \times \int_{0}^{\infty}\frac{\textrm{e}^{-Dk^{2}t}F_{m}\left(k,r\right)F_{m}\left(k,r_{0}\right)}{J_{m}^{2}\left(ka\right)+Y_{m}^{2}\left(ka\right)}kdk
\end{align}
with
\begin{equation}
F_{m}\left(k,r\right) = Y_{m}\left(k r\right)J_{m}\left(k a\right) - J_{m}\left(k r\right)Y_{m}\left(k a\right)
\end{equation}
and the closure relation
\begin{equation}
\int_{0}^{\infty}\frac{F_{m}\left(k,r\right)F_{m}\left(k,r_{0}\right)}{J_{m}^{2}\left(ka\right)+Y_{m}^{2}\left(ka\right)}kdk = \frac{\delta\left(r-r_{0}\right)}{r_0 }
\end{equation}
for any value of $m$ (integer).

The instantaneous flux at a specific point $\left(r=a,\theta\right)$
of the absorbing inner boundary is 
\begin{equation}
j^{*}\left(\theta,t\right)=\frac{D}{\pi^{2}}\sum_{m=-\infty}^{+\infty}\textrm{cos}\left(m\theta\right)\int_{0}^{\infty}\frac{\textrm{e}^{-Dk^{2}t}F_{m}\left(k,r_{0}\right)}{J_{m}^{2}\left(ka\right)+Y_{m}^{2}\left(ka\right)}kdk
\end{equation}
since $F'\left(k,a\right) = \frac{2}{\pi a }$.
The cumulative unbounded flux then reads
\begin{equation}
J^{*}\left(\theta,t\right)=\sum_{m=-\infty}^{+\infty}\frac{\textrm{cos}\left(m\theta\right)}{\pi^{2}}\int_{0}^{\infty}\frac{\left(1-\textrm{e}^{-Dk^{2}t}\right)F_{m}\left(k,r_{0}\right)}{J_{m}^{2}\left(ka\right)+Y_{m}^{2}\left(ka\right)}  \frac{dk}{k }
\end{equation}
and converges to the usual expression
\begin{equation}
\tilde{J}^{*} \left(\theta \right) = \frac{1}{2\pi}\frac{1-\left(a/r_{0}\right)^{2}}{1-2\left(a/r_{0}\right)\textrm{cos}\left(\theta\right)+\left(a/r_{0}\right)^{2}}
\end{equation}
for $t\longrightarrow\infty$, the integral of which over $\theta$
giving $1$ as expected.

\section{\label{app:three}Three dimensions}

The concentration $c\left(r,\theta,t\right)$, $a\le r\le b$, $0\le \theta \le \pi$, where $r$ is the radial distance from the center of the target and $\theta$ is the {\it polar} angle (the coordinate system is chosen so that it does not depend on the azimuthal angle $\phi$) obeys the following diffusion equation 
\begin{align}
\frac{\partial c}{\partial t'}-\frac{1}{r'^{2}}\frac{\partial}{\partial r'} & \left(r'^{2}\frac{\partial c}{\partial r'}\right)  - \frac{1}{r'^{2}\sin\theta}\frac{\partial}{\partial\theta}\left(\sin\theta\frac{\partial c}{\partial\theta}\right)  \nonumber \\
& =\frac{\delta\left(r'-r'_{0}\right)\delta\left(\theta\right)\delta\left(t'\right)}{2\pi r'^{2}\sin\theta}
\end{align}
which has been written in terms of the dimensionless variables $r'=r/a$ and $\tau=a^{2}/D$. The boundary conditions are $c(r=a,\theta,t)=0$, $\partial c/\partial r|_{r=b}=0$. In the remainder, we drop the primes for simplicity.

As in 2D, the solution is symmetric in $\theta$ and takes the usual form 
\begin{equation}
c\left(r,\theta,t\right)=\sum_{m=0}^{+\infty}\left(m+\frac{1}{2}\right)P_{m}\left(\cos\theta\right)\,f_{m}\left(r,t\right)
\end{equation}
with $P_{m}\left(u\right)$ the Legendre polynomial of degree $m$.

The Laplace transform of $f_{m}$ obeys the differential equation 
\begin{equation}\label{eq:diffeqfm3D}
\frac{d^{2}\widehat{f}_{m}}{dx^{2}}+\frac{2}{x}\frac{d\widehat{f}_{m}}{dx}-\left(1+\frac{m\left(m+1\right)}{x^{2}}\right)\widehat{f}_{m}=-\frac{\delta\left(x-x_{0}\right)}{2\pi x_{0}r_{0}}
\end{equation}
where the reduced variable $x=r\sqrt{s}$ has been introduced as previously.

The boundary conditions 
\begin{equation}
\widehat{f}_{m}\left(x_{a}\right)=0\qquad {\rm and}\qquad\frac{\partial\widehat{f}_{m}}{\partial x}\left(x_{b}\right)=0
\end{equation}
imply that the global solution can be written as 
\begin{align}
\widehat{f}_{m<}\left(x\right) & \underset{x<x_{0}}{=}A_{m}\left(i_{m}\left(x\right)k_{m}\left(x_{a}\right)-k_{m}\left(x\right)i_{m}\left(x_{a}\right)\right) \nonumber\\
\widehat{f}_{m>}\left(x\right) & \underset{x>x_{0}}{=}B_{m}\left(i_{m}\left(x\right)k_{m}'\left(x_{b}\right)-k_{m}\left(x\right)i_{m}'\left(x_{b}\right)\right)
\end{align}
where $i_{m}\left(x\right)$ and $k_{m}\left(x\right)$ are the
modified spherical Bessel functions of the first and second
kind of order $m$.

Then, by continuity of $\widehat{f}_{m}$ and integration of Eq. (\ref{eq:diffeqfm3D})  across
the discontinuity in $x_{0}$, one finally gets 
\begin{equation}\label{eq:3Dfmhat}
\widehat{f}_{m}\left(x\right)=\frac{\sqrt{s}}{2\pi}\frac{\left(i_{mb}'k_{m>}-k_{mb}'i_{m>}\right)\left(i_{m<}k_{ma}-k_{m<}i_{ma}\right)}{i_{mb}'k_{ma}-k_{mb}'i_{ma}}
\end{equation}
with $x_{<}=\textrm{min}\left(x,x_{0}\right)$ and $x_{>}=\textrm{max}\left(x,x_{0}\right)$,
and where, for example,
$i_{mb}'$ and $k_{m>}$ stand for $\frac{\partial i_{m}}{\partial x}\left(x_{b}\right)$
and $k_{m}\left(x_{>}\right)$.

The inverse Laplace transform and residue theorem provide the global
solution in terms of dimensional variables 
\begin{align}
c\left(r,\theta,t\right) & =\sum_{m=0}^{+\infty} \frac{\left(m+\frac{1}{2}\right) P_{m}\left(\textrm{cos}\left(\theta\right)\right)}{2\pi i}\int_{\gamma-i\infty}^{\gamma+i\infty}ds\,\textrm{e}^{Dst}\,\widehat{f}_{m}\left(x\right)\nonumber \\
 & =\sum_{m=0}^{+\infty}\left(m+\frac{1}{2}\right)P_{m}\left(\textrm{cos}\left(\theta\right)\right)\,\sum_{n}\textrm{Res}\left[\textrm{e}^{Dst}\,\widehat{f}_{m}\left(x\right)\right]
\end{align}
where the sum of the residues is over the poles of $\widehat{f}_{m}$,
i.e. over the roots $s_{n}$ of the auxiliary equation 
\begin{equation}\label{eq:3Dgs}
g\left(s\right)=i_{m}'\left(qb\right)k_{m}\left(qa\right)-k_{m}'\left(qb\right)i_{m}\left(qa\right)=0
\end{equation}
where $q=\sqrt{s}$.\\

The term $g'\left(s_{n}\right)$ appearing in the residue is 
\begin{equation}
g'\left(s_{n}\right)=\frac{b}{2q_{n}}\left(i_{mb}''k_{ma}-k_{mb}''i_{ma}\right)+\frac{a}{2q_{n}}\left(i_{mb}'k_{ma}'-k_{mb}'i_{ma}'\right)
\end{equation}
(in shorthand notation here, but evaluated in $s=s_{n}$), which can be
simplified thanks to (\ref{eq:3Dgs}) as 
\begin{align}
g'\left(s_{n}\right)&=\frac{b}{2q_{n}}\frac{i_{ma}}{i_{mb}'}\left(i_{mb}''k_{mb}'-k_{mb}''i_{mb}'\right) \nonumber\\
&+\frac{a}{2q_{n}}\frac{i_{mb}'}{i_{ma}}\left(i_{ma}k_{ma}'-k_{ma}i_{ma}'\right)
\end{align}
Then, using the spherical Bessel identities 
\begin{equation}\label{eq:SphericalBesselIdentity}
i_{m}\left(q_{n}a\right)k_{m}'\left(q_{n}a\right)-k_{m}\left(q_{n}a\right)i_{m}'\left(q_{n}a\right)=-\frac{1}{\left(q_{n}a\right)^{2}}
\end{equation}
\begin{equation}
i_{m}''\left(q_{n}b\right)k_{m}'\left(q_{n}b\right)-k_{m}''\left(q_{n}b\right)i_{m}'\left(q_{n}b\right)=-\frac{\left(q_{n}b\right)^{2} + m\left(m+1\right)}{\left(q_{n}b\right)^{4}}
\end{equation}
the function $g'\left(s_{n}\right)$ reads 
\begin{equation}
g'\left(s_{n}\right)=-\frac{1}{2q_{n}^{3}a}\left(\frac{i_{mb}'}{i_{ma}}\right)-\frac{1}{2q_{n}^{3}b}\left(\frac{i_{ma}}{i_{mb}'}\right)\left(1+\frac{m\left(m+1\right)}{\left(q_{n}b\right)^{2}}\right)
\end{equation}

Replacing in (\ref{eq:3Dfmhat}) $k_{mb}'$ by $i_{mb}'k_{ma}/i_{ma}$ thanks to
(\ref{eq:3Dgs}), one obtains 
\begin{align}
c\left(r,\theta,t\right) & =\sum_{m=0}^{+\infty}\left(m+\frac{1}{2}\right)P_{m}\left(\textrm{cos}\left(\theta\right)\right)\,\sum_{n}\textrm{e}^{Dq_{n}^{2}t}\,\frac{aq_{n}^{4}}{\pi}\nonumber \\
 & \times \frac{\left(k_{m>}i_{ma}-i_{m>}k_{ma}\right)\left(k_{m<}i_{ma}-i_{m<}k_{ma}\right)}{1+\frac{a}{b}\left(\frac{i_{ma}}{i_{mb}'}\right)^{2}\left(1+\frac{m\left(m+1\right)}{\left(q_{n}b\right)^{2}}\right)}
\end{align}
where the product $\left(k_{m>}i_{ma}-i_{m>}k_{ma}\right)$ $\left(k_{m<}i_{ma}-i_{m<}k_{ma}\right)$
may also be written as $\left(k_{mr}i_{ma}-i_{mr}k_{ma}\right)\left(k_{mr_{0}}i_{ma}-i_{mr_{0}}k_{ma}\right)$
in terms of the initial variables $r$ and $r_{0}$.\\

Since all the solutions $q_{n}=\sqrt{s_{n}}$ of $g\left(s\right)=0$
are purely imaginary: $q_{n}=i\,k_{n}$, with $k_{n}\in\mathbb{R}$,
one can switch to the spherical Bessel functions of the first and
second kind, $j_{m}$ and $y_{m}$, using the relations 
\begin{align}
i_{m}\left(ik_{n}u\right) & =i^{m}j_{m}\left(k_{n}u\right) \nonumber \\
i_{m}'\left(ik_{n}u\right) & =i^{m-1}j_{m}'\left(k_{n}u\right) \\ 
k_{m}\left(ik_{n}u\right) & =-i^{-m}\left[j_{m}\left(k_{n}u\right)-iy_{m}\left(k_{n}u\right)\right] \nonumber
\end{align}

The solution eventually takes the form 
\begin{align}
c\left(r,\theta,t\right) & =\sum_{m=0}^{+\infty}\left(m+\frac{1}{2}\right)P_{m}\left(\textrm{cos} \; \theta \right) \nonumber \\
& \times  \sum_{n=1}^{+\infty}\textrm{e}^{-Dk_{mn}^{2}t}\,\frac{ak_{mn}^{4}}{\pi}
\frac{f_{mn}\left(r\right)f_{mn}\left(r_{0}\right)}{c_{mn}}
\end{align}
with 
\begin{equation}
f_{mn}\left(r\right)=y_{m}\left(k_{mn}r\right)j_{m}\left(k_{mn}a\right)-j_{m}\left(k_{mn}r\right)y_{m}\left(k_{mn}a\right)
\end{equation}
and
\begin{equation}
c_{mn} = \frac{a}{b}\left(\frac{j_{m}\left(k_{mn}a\right)}{j_{m}'\left(k_{mn}b\right)}\right)^{2}\left(1-\frac{m\left(m+1\right)}{\left(k_{mn}b\right)^{2}}\right)-1
\end{equation}
and where the $k_{mn}$ value is the $n^{th}$ solution of the $m^{th}$
auxiliary equation 
\begin{equation}\label{eq:rootk3D}
j_{m}'\left(kb\right)y_{m}\left(ka\right)-y_{m}'\left(kb\right)j_{m}\left(ka\right)=0
\end{equation}

In 3D, the completeness relation of the basis functions $k_{mn}^2 f_{mn}\left(r\right) / \sqrt{c_{mn}}$ reads (whatever the value of $m$)
\begin{align}
 \frac{2a}{\pi} \, \sum_{n=1}^{+\infty} {k_{mn}^{4}}
\frac{f_{mn}\left(r\right)f_{mn}\left(r_{0}\right)}{c_{mn}} = \frac{\delta\left(r-r_{0}\right)}{r  r_0 } 
\end{align}
while $\sum_{m=0}^{+\infty}\left(m+\frac{1}{2}\right)P_{m}\left(\textrm{cos} \; \theta \right) = \delta \left(\theta\right) / \textrm{sin} \; \theta $.

The instantaneous flux on a specific spherical strip situated at $\left(r=a,\theta\right)$
on the absorbing inner boundary is 
\begin{align}\label{eq:flux3D}
j_{3D}\left(\theta,t\right) & =\left. 2 \pi a^2 D\frac{\partial c}{\partial r}\right| _{a} \nonumber \\
 & =2 a D\sum_{m=0}^{+\infty}\left(m+\frac{1}{2}\right)P_{m}\left(\textrm{cos} \; \theta \right) \nonumber \\
 & \times \sum_{n=1}^{+\infty} \frac{k_{mn}^{3} f_{mn}\left(r_{0}\right)\textrm{e}^{-Dk_{mn}^{2}t}}{c_{mn}} 
\end{align}
since $\frac{\partial f_{mn}}{\partial r} \left( a \right) = 1/k_{mn} a^2$ thanks to the spherical Bessel identity (\ref{eq:SphericalBesselIdentity}).

The cumulative hitting probability $J_{3D}\left(\theta,t\right)=\int_{0}^{t}j_{3D}\left(\theta,t'\right)dt'$ in $\theta$ at time $t$ is
\begin{align}
J_{3D}\left(\theta,t\right)  & =2 a \sum_{m=0}^{+\infty}\left(m+\frac{1}{2}\right)P_{m}\left(\textrm{cos} \; \theta \right) \nonumber \\
 & \times \sum_{n=1}^{+\infty} \frac{k_{mn} f_{mn}\left(r_{0}\right) \left( 1-\textrm{e}^{-Dk_{mn}^{2}t} \right)} {c_{mn}}
\end{align}
whose integral over $\theta$ converges monotonically to $1$ for $t\rightarrow\infty$ (whatever the value of $r_0$):
\begin{equation}
\int_{0}^{\pi}J_{3D}\left(\theta,t \rightarrow \infty \right) \sin\theta d\theta = 2 a \sum_{n=1}^{+\infty}\frac{k_{0n} f_{0n}\left(r_{0}\right)}{c_{0n}} = 1
\end{equation}

The total flux to the inner absorbing sphere can also be computed:
\begin{align}
j_{3D,tot}\left(t\right) & = \int_{0}^{\pi}j_{3D}\left(\theta,t\right) \sin\theta d\theta  \nonumber \\
& = 2 a D \sum_{n=1}^{+\infty}\frac{f_{0n}\left(r_{0}\right) k_{0n}^{3} \, \textrm{e}^{-Dk_{0n}^{2}t}}{c_{0n}}
\end{align}

The eventual hitting probability distribution, $\tilde{J}_{3D}\left(\theta\right)$,
can be obtained directly  from the Laplace transform of $c\left(r,\theta,t\right)$:
\begin{align}
\tilde{J}_{3D}\left(\theta\right) & =\left.2\pi a^{2}D\int_{0}^{\infty}dt\;\frac{\partial c}{\partial r}\right|_{r=a} \\
& =2\pi a^{2}D\frac{\partial}{\partial r}\widehat{c}\left(r,\theta,s\right)\mid_{r=a,s\rightarrow0}\nonumber \\
& \left.=\sum_{m=0}^{+\infty}\left(m+\frac{1}{2}\right)P_{m}\left(\cos\theta\right)\,\left(\frac{i_{m0}k_{mb}'-k_{m0}i_{mb}'}{i_{ma}k_{mb}'-k_{ma}i_{mb}'}\right)\right|_{s\rightarrow0}\nonumber
\end{align}
which gives
\begin{align}\label{eq:finalsolutionsum3D}
\tilde{J}_{3D}\left(\theta\right)&=\sum_{m=0}^{+\infty}\left(m+\frac{1}{2}\right)P_{m}\left(\cos\theta\right) \; \left(\frac{a}{r_{0}}\right)^{m+1} \nonumber \\
& \times \frac{\left(m+1\right)r_{0}^{2m+1}+mb^{2m+1}}{\left(m+1\right)a^{2m+1}+mb^{2m+1}} \\
& > {\tilde{J}^*}_{3D}\left(\theta\right) \nonumber 
\end{align}
in the limit of an infinitely distant wall.

\bigskip

The eventual hitting probability $\tilde{J}_{3D}$ in the 3D confined system is greater than in the unbounded system, which is is in stark contrast to the 2D case, and also appears in dimensions larger than 3.

\bibliography{search2a-25octobre2019}

\end{document}